\documentclass[aps,nofootinbib,showpacs,twocolumn,longbibliography]{revtex4-1}
\usepackage{amstext,amsmath,amssymb,amsfonts,bbm}
\usepackage[latin1]{inputenc}
\usepackage{graphicx}
\usepackage{epstopdf}
\usepackage{amsthm}
\usepackage{tocvsec2}
\usepackage{enumerate}
\usepackage{subfigure}
\usepackage{color}

\newcommand{\lyxdot}{.}
\usepackage{multirow} \usepackage{rotating}
\usepackage[colorlinks,citecolor=blue,linkcolor=blue,urlcolor=blue]{hyperref}

\def\beq{\begin{equation}}
\def\be{\begin{equation}}
\def\ee{\end{equation}}
\def\bes{\begin{eqnarray}}
\def\ees{\end{eqnarray}}

\begin{document}

\title{Statistical discrete geometry}

\author{S. Ariwahjoedi$^{1,3}$, V. Astuti$^{1}$, J. S. Kosasih$^{3}$, C. Rovelli$^{1,2}$, F. P. Zen$^{3}$\vspace{1mm}}

\affiliation{$^{1}$Aix Marseille Universit\'e, CNRS, CPT, UMR 7332, 13288 Marseille, France.\\
$^{2}$Universit\'e de Toulon, CNRS, CPT, UMR 7332, 83957 La Garde, France.\\$^{3}$Institut Teknologi Bandung, Bandung 40132, West Java, Indonesia.}

\begin{abstract} 

\noindent
Following our earlier work \cite{key-4.3}, we construct statistical
discrete geometry by applying statistical mechanics to discrete (Regge)
gravity. We propose a coarse-graining method for discrete geometry
under the assumptions of atomism and background independence. To maintain
these assumptions, we propose restrictions to the theory by introducing
cut-offs, both in ultraviolet and infrared regime. Having a well-defined
statistical picture of discrete Regge geometry, we take the infinite
degrees of freedom (large $n$) limit. We argue that the correct limit
consistent with the restrictions and the background independence concept
is not the continuum limit of statistical mechanics, but the thermodynamical
limit.
\end{abstract}

\maketitle

\section{Introduction}

Attempts to understand the thermodynamical aspect of general relativity
(GR) have already been studied through the thermodynamics of black
holes \cite{key-1.1,key-1.2,key-1.3,key-8.5}. The first step to understand
this problem is the discovery of the four laws of black hole mechanics,
which resembles the laws of thermodynamics \cite{key-1.4}. According
to the laws, black hole could emit  Hawking radiation, due to the
quantum effect of the infalling particles near the black hole's horizon
\cite{key-1.5,key-1.6}. As a consequence of the radiation, black
holes must have temperature and entropy \cite{key-1.5,key-1.6,key-1.7,key-1.8}.
The entropy should be a result of the existence of microstates of
the black hole, which is not believed to exist before, due to the
no-hair theorem \cite{key-1.9,key-1.10,key-1.11}. Loop quantum gravity
(LQG) provides an explanation to the origin of Bekenstein-Hawking
entropy from finite microstates-counting of the black hole \cite{key-1.1,key-1.2}.
The latest results regarding the entropy of black hole from LQG are
presented \cite{key-8.5,key-1.12}.

Attempts to quantize gravity have also developed, with canonical loop
quantum gravity as one of these theories \cite{key-3.1,key-3.2,key-3.30,key-3.31,key-3.38}.
The successful theory of quantum gravity should give a 'correct' classical
theory, -general relativity, in an appropriate limit \cite{key-8.1,key-1.13,key-1.14}.
In loop quantum gravity, there are two parameters which can be adjusted
to obtain this limit: the \textit{spin-number} $j,$ which describes
the size of the quanta of space, and the \textit{number of quanta}
$n,$ which describes the degrees of freedom in the theory. General
relativity is expected to be obtained by taking the \textit{classical
limit}: the limit of large number of degrees of freedom and spins,
$n,j\rightarrow\infty$; while the \textit{mesoscopic}, or the \textit{semi-classical
limit} is obtained by taking only the spin number $j$ to be large,
$j\rightarrow\infty$ \cite{key-8.1,key-1.13,key-1.14}. Tullio Regge
have shown that discrete gravity will coincide with GR in the classical
limit, at the level of the action, when the discrete manifold converge
to Riemannian manifold \cite{key-3.19,key-3.23}. Moreover, Roberts
have shown that the $6j$-symbol, which is used to describe the transition
amplitude in canonical LQG \cite{key-3.31,key-3.20}, in the large-$j$
limit can be written as some functions, such that the transition amplitude
of LQG coincides with the partition function of Regge discrete geometry
\cite{key-7.21}. Attempts to obtain the continuum and semi-classical
limit have already been studied \cite{key-8.1,key-7.15,key-1.15,key-1.16},
with the latest result is reported elsewhere \cite{key-1.13}.

This work is particularly more focused on the large $n$ limit. To
handle a system with large degrees of freedom, it is convenient to
use statistical mechanics as a coarse-graining tool. On the other
hand, thermodynamics can also be explained from statistical mechanics
point of view, as an 'effective' theory emerging from the large degrees
of freedom limit. Therefore, by using the language of statistical
mechanics, the two distinct problems of gravity, namely, its thermodynamical
aspects and the correct classical limit of quantum gravity, appear
to be related to each other.

In this article, we construct statistical discrete geometry by applying
statistical mechanics formulation to discrete (Regge) geometry. Our
approach is simple: by taking the foliation of spacetime (compact
space foliation in the beginning, then the extension to a non-compact
case is possible) as our system, dividing it into several partitions,
similar with many-body problem in classical mechanics, and then characterizing
the whole system using fewer partitions, coarser degrees of freedom.
We propose a \textit{coarse-graining} procedure for discrete geometry
using the lengths, areas, and volumes of the space as the variables
to be taken the coarser, average value. 

Two basic philosophical assumptions are taken in this work. These
assumptions are fundamental in non-perturbative quantum gravity approach
\cite{key-3.1,key-3.2,key-3.31}. First, is the \textit{atomism} philosophy,
which is the mainstream point of view adopted by physics in this century
\cite{key-12,key-13}; it assumes that every existing physical entities
must be countable, which could be a humble way to treat infinities.
Second, is the \textit{background independence} concept \cite{key-3.2,key-3.34,key-3.35},
a point of view shared by many conservative quantum gravity theories,
such as LQG, causal dynamical triangulation, and others. Loosely speaking,
it assumes that the atoms of space do not need to live in another
'background' space, so that we could localize these atoms. In other
words, these atoms of space \textit{are} the space, and we should
not assume any other 'background stage' for them.

To maintain the atomism philosophy, we propose two restrictions in
our procedure by introducing \textit{cut-offs} for the degrees of
freedom, both in ultraviolet and infrared regions. In discrete geometry,
these cut-offs truncate the theory with infinite degrees of freedom
to a theory with finite degrees of freedom. As a direct consequence
to these restriction, we obtain bounds on informational entropy, with
consistent physical interpretations in the classical case.

A clear, well-defined procedure of a coarse-graining method will provide
a first step to understand the problem mentioned in the beginning:
the thermodynamical properties of general relativity and the correct
classical limit of quantum gravity. In the last chapter of this work,
we discuss the possibility to obtain the classical 'continuum' limit
of the kinematical LQG. Using the inverse method of coarse-graining
called as \textit{refinement}, and taking the limit of infinite degrees
of freedom, we argue that the correct limit consistent with our two
basic philosophical assumptions is \textit{not} the continuum limit
of statistical mechanics, but the thermodynamical limit. If this thermodynamical
limit exist, theoretically, we could obtain the equation of states
of classical gravity from (Regge) discrete geometry for the classical
case.

The article contains four sections. In Section II, we propose a clear,
well-defined coarse-graining procedure for any arbitrary classical
system, and introduce two types of restriction to finitize the informational
entropy. In Section III, we apply our coarse-graining procedure to
discrete geometry, say, a set of connected simplices. A new results
obtained in this chapter is the proposal of well-defined coarse-graining
method and statistical discrete geometry formulation. The last section
is the conclusion, where we give sketches about the generalization
to the non-compact space case. We argue that the correct limit which
will bring us to classical general relativity is not the 'continuum'
limit, but the thermodynamical limit of the statistical mechanics.

\section{Classical coarse-graining proposal}

\subsection{Definition of coarse-graining}

In this section, we propose a precise definition of coarse-graining
in the classical framework. It is defined as a procedure containing
two main steps: \textbf{reducing the degrees of freedom away to obtain
reduced-microstates} and then, \textbf{taking the average state of
all these possible reduced-microstates}. A specific example is used
to help to clarify the understanding. Suppose we have a classical
system with two degrees of freedom. Let the degrees of freedom be
represented by the energy of each partition $\left\{ E_{1},E_{2}\right\} $,
and let the coarse-graining procedure be defined as a way to represent
this system using a single degrees of freedom.

\subsubsection{Center-of-mass variables}

Our system is characterized by variables $\left\{ E_{1},E_{2}\right\} $,
with $E_{1}$ and $E_{2}$ represent the energy of the first and second
partitions, respectively. Suppose there exists a canonical 'transformation'
such that:
\begin{eqnarray}
E_{\pm} & = & E_{1}\pm E_{2},\label{eq:3.}
\end{eqnarray}
so that we have a new canonical variables $\left\{ E_{+},E_{-}\right\} $,
usually called as the \textit{center-of-mass} variables. Each individual
variable describes different thing: $\left\{ E_{1},E_{2}\right\} $
describe the energy of each partition, while $\left\{ E_{+},E_{-}\right\} $
describe the total energy of the system, and the energy difference
between these two partitions. Nevertheless, both $\left\{ E_{1},E_{2}\right\} $
and $\left\{ E_{+},E_{-}\right\} $ describe the same system and have
same amount of informations.

The next step is to treat this binary system as a single system, described
by a single degree of freedom. This can be done naturally by taking
only $E_{+}$ and neglecting the information of $E_{-}$. But having
only $E_{+},$ the correct values of $\left\{ E_{1},E_{2}\right\} $
can not be obtained, because there are no enough informations. With
the correct $E_{-}$ unknown, then we obtain a set $\gamma$ representing
$E_{+,}$ with all possible values of $E_{-}$.

\subsubsection{Microstates, macrostate, sample space, and classical coarse-graining}

In this subsection, we define new terminologies, which is slightly
different with the usual statistical mechanics terminologies. Let
$\mathbb{\gamma}$ be the \textit{physical configuration space}. Let
any point $\left\{ E_{1}^{\left(j\right)},E_{2}^{\left(j\right)}\right\} $
or $\left\{ E_{+}^{\left(j\right)},E_{-}^{\left(j\right)}\right\} $
element of $\mathbb{\gamma}$ be the \textit{$j^{th}$-microstates}%
\footnote{In a precise terminology, the microstates should contain the generalized
momenta element, but we just neglect it since we are not going to
use it in this section%
}. Then we define the (total) energy of the microstate as $E_{1}^{\left(j\right)}+E_{2}^{\left(j\right)},$
which is exactly $E_{+}^{\left(j\right)}.$ For the discrete case,
the microstate is $\left\{ E_{1}^{\left(j\right)},E_{2}^{\left(j\right)}\right\} $
or $\left\{ E_{+}^{\left(j\right)},E_{-}^{\left(j\right)}\right\} ,$
an element of a discrete set $\mathbb{\gamma}$ and $j\in\mathbb{Z},$
while for the continuous case, it is $\left\{ E_{1}\left(x_{1}\right),E_{2}\left(x_{2}\right)\right\} $
or $\left\{ E_{+}\left(x_{1},x_{2}\right),E_{-}\left(x_{1},x_{2}\right)\right\} ,$
an element of continuous set $\mathbb{\gamma}$ and $x_{1},x_{2}\in\mathbb{R}.$
Let $\left\{ E_{1},E_{2}\right\} $ or $\left\{ E_{+},E_{-}\right\} $
be the \textit{correct microstates} and $E_{1}+E_{2}=E_{+}$ as the
'correct' total energy. 

Finally, we define the \textit{reduced $j^{th}$-microstate} as $E_{+}^{\left(j\right)}$
and the \textit{correct reduced microstate} as the 'correct' total
energy $E_{+}$. Our\textit{ macrostate} is the \textbf{average over
all possible reduced-microstates}, under a \textit{probability distribution
/ weight} $p_{j}$ for the discrete case, or a distribution function
$p\left(\mathbf{x}\right)$ for the continuous case:

\begin{eqnarray}
\overline{E} & = & \sum_{j}p_{j}E_{+}^{\left(j\right)},\qquad\textrm{(discrete case),}\label{eq:3.0}\\
\overline{E} & = & \intop_{\gamma}d\mathbf{x}\; p\left(\mathbf{x}\right)E_{+}\left(\mathbf{x}\right),\qquad\textrm{(continuous case),}\label{eq:3.0aa}
\end{eqnarray}
with $\mathbf{x}=\left(x_{1},x_{2}\right)$ is the measure on $\gamma$.
The \textit{sample space} $\mathbb{X}$ is the \textbf{space of all
possible microstates}, which in this case, is $\gamma$. We define
the macrostate $\bar{E}$ as our \textit{coarse-grained variable},
the microstates as the \textit{fine-grained variables}, and the procedure
to obtain the macrostate as \textit{coarse-graining}.

\subsection{Restrictions}

\subsubsection{Distribution function and informational entropy}

A specific probability distribution $p_{i}$ gives the probability
for a specific microstate $i$ to occur. It defines a related entropy
for the system. The definition of entropy we use in our work is the
\textit{informational entropy} or \textit{Shannon entropy}, defined
mathematically as \cite{key-2.2}: 
\begin{equation}
S=-\sum_{i}p_{i}\ln p_{i}.\label{eq:3.0a}
\end{equation}
Moreover, we will use the generalization of Shannon entropy, the \textit{differential
entropy} \cite{key-2.3}: 
\begin{equation}
S=-\intop_{0}^{\infty}\mathbf{dx}\: p\left(\mathbf{x}\right)\ln p\left(\mathbf{x}\right),\label{eq:3.0b}
\end{equation}
for a given continuous probability density $p\left(\mathbf{x}\right)\in C^{m}\left[\mathbb{R}^{n}\right]$,
which is just the continuous version of Shannon entropy. 

The informational entropy is the amount of lack-of-information of
the coarse-grained state, with respect to the fine-grained state.
See an analog-example in FIG. 1. 

\begin{figure}[h]
\centerline{\includegraphics[height=3.5cm]{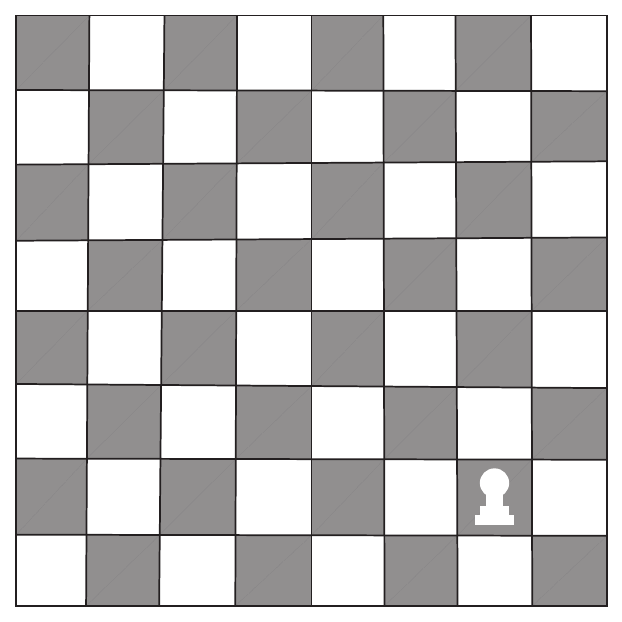}} \caption[A nice analogy to help in understanding the informational entropy.]{A nice analogy to help in understanding the informational entropy
is by using a chess-board game. Suppose we want to know the location
of the white pawn on the board. To know this, we need to ask questions
such that the answers are only 'yes' or 'no'. We can start by asking
with: \textit{'Is the pawn located at the half-left of the board?}',
which will follow with the answer 'no'. We can repeat such question
until we arrive at the right square. At this point, our information
about the position of the pawn is already complete, and the entropy
of the system is zero, since there is no lack-of-information in a
'completely-known' system with maximal certainty. }
\end{figure}

We adapt the analogy to our microstates. The chess-board will be the
sample space $\mathbb{X}$: the space of all possible microstates
which correspond to a specific macrostate $\bar{E}$. Our 'correct'
microstate is a single point on the sample space $\mathbb{X}$, which
is the location of the pawn in the example in FIG. 1. To arrive at
this correct microstate, we need to ask question: \textit{'Is the
correct microstate located at some region $\mathbb{X}'\subset\mathbb{X}$?}'.
If we could answer all the questions with binary answers (\textit{i.e.},
with 'yes / no' answer) until we arrive to the correct microstate,
then we know \textit{completely} the information about the fine-grained
state. This is the same as having the \textbf{Dirac-delta distribution}
as a distribution function: 
\begin{equation}
p\left(\mathbf{x}\right)=\delta\left(\overline{\mathbf{x}}-\mathbf{x}\right),\label{eq:3.0c}
\end{equation}
satisfying:\textbf{ 
\[
\intop_{0}^{\infty}\mathbf{dx}\:\delta\left(\overline{\mathbf{x}}-\mathbf{x}\right)f(\mathbf{x})=f(\overline{\mathbf{x}}),\quad\forall f,
\]
}with the specification $f(\overline{\mathbf{x}})=0$ if $\overline{\mathbf{x}}$
is not in the support of $f$. The distribution $\delta\left(\overline{\mathbf{x}}-\mathbf{x}\right)$
is normalized to unity: 
\[
\intop_{0}^{\infty}\mathbf{dx}\:\delta\left(\overline{\mathbf{x}}-\mathbf{x}\right)=1.
\]

The corresponding informational entropy of a completely-known system
should be zero. Obviously, the distribution function of a completely-known
state should be the Dirac-delta function, since it points directly
to a single microstate parametrized by $\overline{\mathbf{x}}$. By
this reasoning, we expect the differential entropy of the Dirac-delta
function is zero, but inserting (\ref{eq:3.0c}) to (\ref{eq:3.0b})
gives: 
\begin{eqnarray}
S=-\intop_{0}^{\infty}\mathbf{dx}\:\delta\left(\overline{\mathbf{x}}-\mathbf{x}\right)\ln\delta\left(\overline{\mathbf{x}}-\mathbf{x}\right) & = & -\infty.\label{eq:3.0d}
\end{eqnarray}
This divergence comes from the fact that we have an infinitesimally
small scale, which in physical term is known as the \textit{ultraviolet
divergence} \cite{key-2.4}.

\subsubsection{Restrictions and bounds on entropy}

In order to maintain the atomism philosophy as our starting point,
we need to give restrictions to the sample space, and this results
to the finiteness of the informational entropy \cite{key-2.5}.

\paragraph{Lower bound entropy}

A continuous system, either finite or infinite, contains infinite
amount of informations, which means it could have an infinite and/or
arbitrarily small entropy. According to the atomism philosophy, it
is necessary to have a \textit{finite} amount of informations given
a finite system, so that we only need to provide finite informations
to know exactly the 'correct' microstate of the system. Intuitively,
we would like to have the entropy of the Dirac-delta function to be
zero, since it points-out a single point on the sample space, which
means we know completely the informations of the correct microstate.
But from the calculation carried in the previous section, the differential
entropy of the Dirac-delta function is $-\infty$. This is a consequence
of allowing an infinitesimally-small scale in the theory, which leads
to the UV-divergence. Therefore, to prevent UV-divergency, we consider
the \textit{discrete} version of the Dirac-delta, the \textbf{Kronecker-delta},
which is defined as: 
\[
\delta_{ij}\left\{ \begin{array}{c}
=0,\quad i\neq j\\
=1,\quad i=j
\end{array}\right.,
\]
normalized to unity: 
\[
\sum_{i=0}^{\infty}\delta_{ij}=1.
\]

Adapting to our case, let the index $i$ runs from zero to any number
(it can be extended to infinity), and let the index $j$ denotes our
correct microstate. Obviously, the Kronecker-delta is the probability
distribution of a completely-known system for the discrete version.
We call the system in such condition as a '\textit{pure} state', following
the statistical quantum mechanics terminology. The Shannon entropy
of this pure state is, by (\ref{eq:3.0a}): 
\[
S_{\Delta}=-\sum_{i}^{\infty}\delta_{ij}\ln\delta_{ij}=0.
\]
Therefore, in order to have a well-defined entropy for a pure state
which is consistent with the physical interpretation, namely, the
completely-known system, we need to give a \textit{restriction} to
the sample space $\mathbb{X}$:

\textbf{Restriction 1 (Strong):} \textit{The sample space must be
discrete, in order to have a well-defined minimum entropy for a completely-known
system with maximal certainty.}

This means the possible microstates must be \textit{countable.} The
physical consequence of our first restriction is that all the corresponding
physical properties (the 'observables' in quantum mechanics term)
must have discrete values. This statement, for our case, is equivalent
with the statement: the spacing of the energy level can not be infinitesimally-small,
there should exist an ultraviolet cut-off which prevent the energy
levels to be arbitrarily small. This is exactly the expectation we
have from the quantum theory.

Since the Shannon entropy (\ref{eq:3.0a}) is definite positive, using
the first restriction, we could define the lower bound of the informational
entropy: 
\[
S_{\textrm{min}}=0,
\]
which comes from the Kronecker-delta probability distribution, describing
a completely-known state. This is in analog with the von-Neumann entropy
bound on the pure state in quantum mechanics \cite{key-2.6}.

\paragraph{Upper-bound entropy}

The completely-known 'pure' state is an ideal condition where we know
completely, without any uncertainty, \textit{all} the informations
of the correct microstate describing the system. This, sometimes,
is an excessive requirement. In general, it is more frequent to have
a state with an incomplete and uncertain information, a coarse-grained
state. The probability distribution of this state could vary, from
the normal to the homogeneous distribution. 

Suppose we already obtain a specific probability distribution for
the system, then we can ask questions: '\textit{How much information
we already know about the state? How certain is our knowledge about
the state? How much information we need to provide to know the correct
microstate?}' The answer to these questions depends on how much entropy
we have when we obtain \textit{least} information (maximal uncertainty)
about the correct microstate.

Let us consider a state where we have no certainty in information
about the correct microstate. Obviously, the probability distribution
of such state should be a homogeneous distribution: 
\begin{equation}
p_{i}=c,\label{eq:3.1}
\end{equation}
with $c$ is a constant, and is normalized to unity: 
\[
\sum_{i}^{W}p_{i}=Wc=1,\qquad c=\frac{1}{W}.
\]
In general, the discrete sample-space $\mathbb{X}$ is characterized
by the index $i$ running from $0$ to $\infty,$ by the first restriction.
If we extend the number of possible microstates $W$ to infinity,
$c$ will go to zero. As a consequence, for an infinite, countable
sample space, the 'upper bound' of the entropy is $\infty$. Therefore,
to obtain the correct microstate for such sample space, we need to
provide an infinite amount of informations. This is not favourable,
since the total amount of informations in the universe should be finite
\cite{key-2.7}. It is necessary to add another restriction.

\textbf{Restriction 2 (Weak)}: \textit{The sample space must be discrete
and finite, in order to have a well-defined maximal entropy for a
completely-unknown state with maximal uncertainty.}

It means the number of possible microstates must be finite, besides
being countable: 
\[
W<\infty.
\]
The physical implication of the second restriction is that all the
properties we want to take average, i.e., the properties of microstates
$\left\{ E_{1},E_{2}\right\} $ and $\left\{ E_{+},E_{-}\right\} $
can \textit{not} be infinitely large.

With the first and second restrictions, we can obtain the entropy
for the homogeneous probability distribution: 
\begin{equation}
S_{\Delta}=-\sum_{i=0}^{W}c\ln c=-\ln\frac{1}{W}=\ln W,\label{eq:9}
\end{equation}
which gives Boltzmann definition of entropy \cite{key-2.8}: ``the
logarithm of number of all possible microstates''. This gives an
upper bound on the entropy: 
\[
S_{\textrm{max}}=\ln W.
\]

In conclusion, with first and second restrictions, we have a well-defined
entropy for any probability distribution, within the range: 
\begin{equation}
S_{\textrm{min}}\leq S\leq S_{\textrm{max}},\label{eq:y}
\end{equation}
with $S_{\textrm{min}}=0$ and $S_{\textrm{max}}$ is the entropy
of a completely 'ordered' (completely-known) and completely random
(completely-unknown) system, respectively.

\subsection{General ensemble case}

The example we used in the previous section is the \textit{microcanonical}
case. The correct microstate is $\left\{ E_{1},E_{2}\right\} $, and
the total energy of the correct microstate is $E_{+}=E_{1}+E_{2},$
which is \textit{also} the correct reduced microstate. The microcanonical
case is defined by the \textit{microcanonical constraint}; this constraint
need to be satisfied by any generic microstates:
\begin{equation}
E_{+}^{\left(j\right)}=E_{1}^{\left(j\right)}+E_{2}^{\left(j\right)}\equiv E_{+},\label{eq:3.3a}
\end{equation}
which means $\left\{ E_{1}^{\left(j\right)},E_{2}^{\left(j\right)}\right\} $
may vary, but their sum must be equal to the correct reduced microstate
$E_{+}.$ All these possible microcanonical microstates describe a
'hypersurface' of constant energy in $\mathbb{R}^{2}$, which is a
straight line $\gamma$.

In standard statistical mechanics, the microcanonical ensemble describe
an isolated system, where the number of partitions and the energy
of the microstates are conserved: $\triangle E=0,$ $\triangle N=0$.
For such  system, we need to made an \textit{a priori} assumption,
which is known as the \textbf{fundamental postulate of statistical
mechanics}: \textit{``All possible microstates of an isolated system
have equal weight / probability distribution}.'' \cite{key-2.9}.
This means the probability distribution $p_{j}$ in (\ref{eq:3.0})
is a constant, as in relation (\ref{eq:3.1}). Moreover, this suggests
that the microcanonical system is a completely-unknown system with
maximal uncertainty. This assumption is very useful, because it is
used to derive the probability distribution for a more general system.

\subsubsection{Canonical case and energy fluctuation}

In the microcanonical case, our system is closed and isolated, hence
we do not have any interaction between the system with the environment.
For the next section, we will release this restriction and use a closed
(not necessarily need to be isolated)  system as an example. First,
we start with the canonical case, where only the number of partitions
is conserved, while the energy can freely flow into and out from the
system: $\triangle E\neq0,$ $\triangle N=0$.

Let our \textit{system} $\mathbb{S}$ be a portion (or the subsystem)
of the \textit{universe} $\mathbb{U}$, $\mathbb{S}\subset\mathbb{U}$.
The universe is, \textit{a priori,} an isolated system, well-defined
by a microcanonical ensemble, with equal weight for all its microstates.
The 'outside' is defined as the complement of $\mathbb{S}$ with respect
to $\mathbb{U}$, say $\bar{\mathbb{S}},$ such that $\mathbb{S}\cup\bar{\mathbb{S}}=\mathbb{U},$
and $\mathbb{S}\cap\bar{\mathbb{S}}=\varnothing.$ We called $\bar{\mathbb{S}}$
as the \textit{environment}, due to the definition in \cite{key-2.10}.
$\mathbb{S}$ and $\bar{\mathbb{S}}$ together, construct a \textit{bipartite}
system, see FIG. 2. 
\begin{figure}[h]
\begin{centering}
\includegraphics{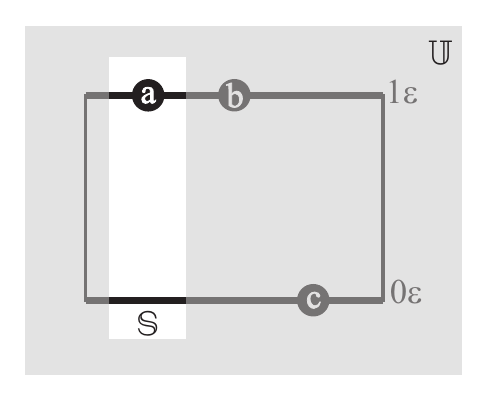}
\par\end{centering}

\centering{}\caption[A bipartite system described by two level of energy.]{A bipartite system described by two level of energy: $\left\{ 0\varepsilon,1\varepsilon\right\} $.
The universe $\mathbb{U}$ consist three particles $\left\{ a,b,c\right\} .$
We choose our system as a subset $\mathbb{S}\subset\mathbb{U}$, which
contains $\left\{ a\right\} .$ The environment is then the complement
$\bar{\mathbb{S}}$ containing $\left\{ b,c\right\} $. }
\end{figure}

In the canonical case, energy can flow between $\mathbb{S}$ and $\bar{\mathbb{S}}$,
which indicate an interaction between them. Hence, the total energy
of $\mathbb{S}$ is not conserved, so in general, we release the microcanonical
constraint (\ref{eq:3.3a}):
\begin{equation}
E_{+}^{\left(j\right)}=\sum_{i=1}^{2}E_{i}^{\left(j\right)}.\label{eq:z}
\end{equation}
 Nevertheless, the microstates $\left\{ E_{1}^{\left(j\right)},E_{2}^{\left(j\right)}\right\} $
still need to give the 'correct' macrostate as a constraint equation:
\[
\overline{E}=\sum_{j}^{n}p_{j}E_{+}^{\left(j\right)},\qquad n<\infty,
\]
with $p_{j}$ is the canonical distribution function. It should be
kept in mind that neither $\mathbb{S}$ nor $\bar{\mathbb{S}}$ is
microcanonical, which means neither have equal weight for their microstates;
$p_{j}\neq$ constant.

Provided that we know \textit{exactly} all the possible microstates
of the universe $\mathbb{U}$, theoretically, we could obtain the
exact canonical probability distribution, given a subsystem $\mathbb{S}\subset\mathbb{U}$.
The (normalized) probability distribution for microstate $j$ is:
\[
p_{j}=\frac{W^{\left(j\right)}}{\sum_{k}W^{\left(k\right)}},
\]
with $\sum_{k}W^{\left(k\right)}$ is the sum of all possible microstates
of system $\mathbb{S}$. 

An important point we want to emphasize is: \textit{knowing all possible
microstates of the (isolated) universe $\mathbb{U}$ (which mean knowing
$\mathbb{X}_{U}$), then choosing a specific subsystem $\mathbb{S}\subset\mathbb{U}$
(which means knowing $\mathbb{X}_{S}$), then we could obtain exactly
the weight / probability distribution $p_{j}^{\left(S\right)}$ for
each distinct microstate inside $\mathbb{X}_{S}$, by using the fundamental
postulate on $\mathbb{X}_{U}$}. This is analog with obtaining a density
matrix from a pure state of the universe by tracing out parts of the
Hilbert space in quantum mechanics \cite{key-2.6}. However, knowing
all the possible microstates of the universe is an excessive requirement,
therefore we simplify the case by taking an appropriate limit. Given
a distribution $p_{j}^{\left(S\right)}$ for our canonical system
$\mathbb{S}$, which could be any well-defined distribution, from
Kronecker-delta to homogen distribution, for a limit where the environment
is much more larger than the observed system, $\bar{\mathbb{S}}\ggg\mathbb{S}$,
the most frequent distribution to occur is \cite{key-2.10}:
\begin{equation}
p_{j}=g^{\left(j\right)}\exp\left(\beta E_{+}^{\left(j\right)}\right).\label{eq:3.4}
\end{equation}
Moreover, for a limit where we have large number of energy level and
particles, $E_{+}\approxeq E_{+}^{\left(j_{max}\right)}.$ Therefore,
if the environment is much more larger than the canonical system,
and the energy levels of the system is very dense with a large number
of particles, we can safely use (\ref{eq:3.4}) as the probability
distribution which is responsible for the macrostate $\bar{E}.$ $\beta$
is the Lagrange multiplier related to the microcanonical constraint,
which can be defined as the\textit{ inverse temperature}. See \cite{key-2.11}
for a detailed derivation.

For the canonical case, each microstates could have different weight,
and this will result in the existence of the most probable reduced-microstate
to occur. In other words, the variance, standard deviation, and fluctuation
of energy in the canonical case, in general, are not zero. But in
the limit of large number of energy levels and particles, the standard
deviation of energy $\sigma_{E}\ll\overline{E},$ therefore, the fluctuation
is, effectively, zero. See \cite{key-2.8,key-2.9,key-2.10,key-2.11}
for standard basic definitions in statistical mechanics.

\subsubsection{Grandcanonical case and particle fluctuation.}

The most general ensemble case in statistical mechanics is the\textit{
grandcanonical case}, where number of partitions and total energy
of the system is not conserved: $\triangle E\neq0,$ $\triangle N\neq0$.
We release the canonical constraint (\ref{eq:z}), which contains
the information of the number of partitions of the system. This means,
we allow the change in the number of partitions, which is two in our
previous binary-partition example in the begining of Subsection II
A: 
\[
E_{+}^{\left(j,n\right)}=\sum_{i=1}^{n}E_{i}^{\left(j\right)},
\]
(see equation (\ref{eq:3.3a}) and then compare them).

Let us define an \textit{open systems} $\mathbb{S}'$ inside $\mathbb{U}$
as the collection of all possible subsets of $\mathbb{U}:$ $\mathbb{S}'=\left\{ s_{i}\left|a\in s_{i};a\in\mathbb{U}\right.\right\} ,$
$s_{i}\in\mathbb{U}$. These subsets $s_{i}$ are naturally all the
possible \textit{grandcanonical microstates} of our open system $\mathbb{S}$.
It is clear that the grandcanonical microstates do not only consist
the canonical microstates, but also \textit{other} canonical microstates
\textit{with different number of particles}. We need to add all of
these collections of canonical microstates with different numbers
of particles to obtain all possible grandcanonical microstates.

An exact grandcanonical distribution can be obtained by counting procedure,
but since it is not really effective to do the calculation this way,
we simplify the case just as the previous canonical case. In the limit
where the environment is much larger than the grandcanonical system,
and the level energy of the system is very dense with a large number
of particles, the grandcanonical probability distribution responsible
for the macrostate $\overline{E}$ is:
\[
P_{\left(j,n\right)}=g^{\left(j\right)}\exp\left(\mu\beta n+\beta E_{+}^{\left(j\right)}\right),
\]
with $\mu$ is another Lagrange multiplier related to the canonical
constraint (fixing the number of particles), defined as the \textit{chemical
potential} \cite{key-2.11}. In the limit of large number of energy
levels and particles, the standard deviation of particle's number
$\sigma_{n}\ll\overline{N},$ therefore, the particle fluctuation
is, again, effectively  zero.

\section{Applying coarse-graining to gravity}

In this section, we apply our coarse-graining proposal described in
Section II to discrete gravity. The main idea proposed by the theory
of general relativity is: \textit{gravity is geometry} \cite{key-4.2,key-4.1},
and therefore, coarse-graining gravity is coarse-graining geometry.
In the first subsection, we will start with coarse-graining a 1-dimensional
segment, to give a clear insight of the procedure. In the second subsection,
we will give restrictions which are related to the cut-offs in the
theory. We carry our procedure in a microcanonical framework, but
we will generalize to general ensembles, this is done in the third
subsection of this part. At last, we will coarse-grain higher dimensional
geometries, which are areas and volumes. The dynamics has not been
included in our work.

\subsection{(1+1) D: Coarse-graining segments}

\subsubsection{Transformation to the center-of-mass coordinate}

For the moment, let us work in $(1+1)$-dimension, where the spatial
slice is a 1-dimensional geometrical object. Let us suppose that we
have a system containing two coupled segments. Written in the coordinate-free
variables, the system has three degrees of freedom: $\left\{ \left|\mathbf{l}_{1}\right|,\left|\mathbf{l}_{2}\right|,\phi_{12}\right\} $
\cite{key-7}. $\left|\mathbf{l}_{1}\right|,\left|\mathbf{l}_{2}\right|$
are the lengths of the two segments, while $\phi_{12}$ is the angle
between them. Following the procedure proposed in Section II, to coarse-grain
the system means treating the system as a single segment containing
only single degrees of freedom. The first step of this coarse-graining
procedure is to obtain the transformation to the 'center-of-mass'
coordinate. Let us return to the non-coordinate-free variables, where
we describe the system in a vectorial way, $\left\{ \mathbf{l}_{1},\mathbf{l}_{2}\right\} $
\cite{key-7}. Written in vectors, the transformation to obtain the
center-of-mass coordinate, $\left\{ \mathbf{l}_{+},\mathbf{l}_{-}\right\} $,
is obvious:
\begin{eqnarray}
\mathbf{l}_{\pm} & = & \mathbf{l}_{1}\pm\mathbf{l}_{2},\label{eq:x}
\end{eqnarray}
These are similar to the transformation (\ref{eq:3.}) defined in
Subsection II A. 

The next step is to obtain the coordinate-free version of $\left\{ \mathbf{l}_{+},\mathbf{l}_{-}\right\} $.
It could be done by taking the norms and the 2D dihedral angle between
these two 1-forms, as defined in \cite{key-7}. The 'center-of-mass'
coordinate-free variables are written as $\left\{ \left|\mathbf{l}_{+}\right|,\left|\mathbf{l}_{-}\right|,\alpha\right\} $,
where the transformation is:
\begin{eqnarray}
\left|\mathbf{l}_{\pm}\right| & = & \sqrt{g\left(\mathbf{l}_{1}\pm\mathbf{l}_{2},\mathbf{l}_{1}\pm\mathbf{l}_{2}\right)},\label{eq:4.1}\\
\cos\alpha & = & \frac{g\left(\mathbf{l}_{1}+\mathbf{l}_{2},\mathbf{l}_{1}-\mathbf{l}_{2}\right)}{\left|\mathbf{l}_{1}+\mathbf{l}_{2},\right|\left|\mathbf{l}_{1}-\mathbf{l}_{2}\right|},\label{eq:4.3}
\end{eqnarray}
$\alpha$ is the 2D angle between segment $\left|\mathbf{l}_{+}\right|$
and $\left|\mathbf{l}_{-}\right|$, and $g$ is a Euclidean metric,
see \cite{key-7} for a detail explanation. In fact, we could directly
define transformation (\ref{eq:4.1})-(\ref{eq:4.3}) as the transformation
to the center-of-mass variables without using the vectorial transformation
(\ref{eq:x}); we use the vectorial form just as a simple way to derive
this transformation.

It should be kept in mind that $\left\{ \left|\mathbf{l}_{1}\right|,\left|\mathbf{l}_{2}\right|,\phi_{12}\right\} $
and $\left\{ \left|\mathbf{l}_{+}\right|,\left|\mathbf{l}_{-}\right|,\alpha\right\} $
describe the same system, but they use different choice of variables:
the first one use the length of each individual segments and the angle
between them, while the later use the total length and their difference,
with the angle between them. 

Now, having the 'center-of-mass' variables written in a coordinate-free
way, the coarse-graining procedure is to be defined straightforwardly.

\subsubsection{Geometrical setting and terminologies}

Following an example in \cite{key-7}, for the moment, we take a \textit{compact},
\textit{simply-connected} slice as the foliation of a discrete spacetime
$\mathcal{M}_{\Delta}$, we call this spatial slice $\mathcal{S}_{\triangle}^{1};$
which is topologically a boundary of an $n$-polygon. Each segment
is a portion of a real line $\left[0,\left|\mathbf{l}_{i}\right|\right]\subset\mathbb{R}_{i}^{1}.$
The \textit{length} of each segment is defined as a scalar $\left|\mathbf{l}_{i}\right|.$
We call each segment of $\mathcal{S}_{\triangle}^{1}$ as a \textit{partition}
of the whole system.

Suppose we have $\mathcal{S}_{\triangle}^{1}$ discretized by six
partitions as in FIG. 3. 
\begin{figure}[h]
\centerline{\includegraphics[scale=0.8]{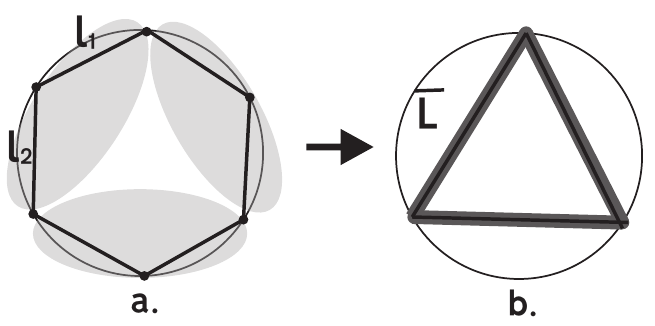}} \caption{(a) is $\mathcal{S}_{\triangle=6}^{1}$ , a hexagon containing by
a six segments. It is coarse-grained into (b), $\mathcal{S}_{\triangle=3}^{1}$
, where it only contains three segments by collecting each two adjacent
segments together, for example, $\left|\mathbf{l}_{1}\right|$ and
$\left|\mathbf{l}_{2}\right|$. Averaging over all the possible values
of $\left|\mathbf{l}_{1}\right|$ and $\left|\mathbf{l}_{2}\right|$
gives the \textit{coarse-grained length} $\overline{\left|\mathbf{L}\right|}$.}
\end{figure}
 In analogy with the coarse-graining of many-body problem described
in Section II, we coarse-grained $\mathcal{S}_{\triangle}^{1}$ so
that it only contains three partitions, by collecting each two adjacent
segments together. We call the discrete geometrical object containing
six and three partitions as a hexagon $\mathcal{S}_{\triangle=6}^{1}$
and triangle $\mathcal{S}_{\triangle=3}^{1}$ , respectively. $\mathcal{S}_{\triangle=6}^{1}$
is the \textit{fine-grained state} and triangle $\mathcal{S}_{\triangle=3}^{1}$
is the \textit{coarse-grained state}. See FIG. 3. 

Let us focus to the two adjacent segments of $\mathcal{S}_{\triangle=6}^{1}$,
they construct a system containing two coupled segments, say, system
$\mathbb{L}$, for 'length'. $\mathbb{L}$ is described by three degrees
of freedom, denoted by the configuration $\left\{ \left|\mathbf{l}_{1}\right|,\left|\mathbf{l}_{2}\right|,\phi_{12}\right\} ,$
or $\left\{ \left|\mathbf{l}_{+}\right|,\left|\mathbf{l}_{-}\right|,\alpha\right\} $,
written in the center-of-mass variables. We call the configuration
$\left\{ \left|\mathbf{l}_{1}\right|,\left|\mathbf{l}_{2}\right|,\phi_{12}\right\} $
or $\left\{ \left|\mathbf{l}_{+}\right|,\left|\mathbf{l}_{-}\right|,\alpha\right\} $
as the \textit{correct microstate} of system $\mathbb{L}$. Any possible
configuration of $\mathbb{L}$, which we denote as $\left\{ \left|\mathbf{l}_{1}^{\left(j\right)}\right|,\left|\mathbf{l}_{2}^{\left(j\right)}\right|,\phi_{12}^{\left(j\right)}\right\} $
or $\left\{ \left|\mathbf{l}_{+}^{\left(j\right)}\right|,\left|\mathbf{l}_{-}^{\left(j\right)}\right|,\alpha^{\left(j\right)}\right\} ,$
for $j\in\mathbb{Z}$, is defined as the \textit{$j^{th}$-microstate}.
The collection of all possible microstates construct a sample space
$\mathbb{X}_{L}$ of the system $\mathbb{L}$. 

Now let us use the center-of-mass variables $\left\{ \left|\mathbf{l}_{+}\right|,\left|\mathbf{l}_{-}\right|,\alpha\right\} $
as the correct-microstate. We want to treat $\mathbb{L}$ as a coarser
system containing \textit{only} a single segment. A natural way to
do this is to describe the system using variable $\left|\mathbf{l}_{+}\right|$,
by transformation (\ref{eq:4.1}). $\left|\mathbf{l}_{+}\right|$
is the \textit{correct reduced-microstate}, which is the \textit{total}
length of the segment: the \textit{norm of the sum} of segment $\mathbf{l}_{1}$
and $\mathbf{l}_{2}$:
\[
\left|\mathbf{l}_{+}\right|=\left|\mathbf{l}_{1}+\mathbf{l}_{2}\right|,
\]
see FIG. 4. 
\begin{figure}[h]
\centerline{\includegraphics[scale=0.8]{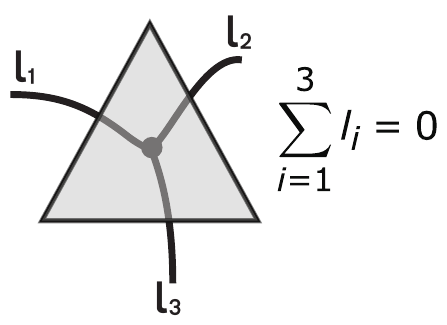}} \caption{A closed triangle in 2-dimensional space is formed by three vectors
$\mathbf{l}_{1}$, $\mathbf{l}_{2}$, $\mathbf{l}_{12}\in T_{p}M\sim\mathbb{R}^{2}$,
which must satisfied a closure condition (sometimes called as Gauss
constraint in physics term) $\mathbf{l}_{1}+\mathbf{l}_{2}+\mathbf{l}_{12}=0$.
Three segments $\left|\mathbf{l}_{1}\right|$, $\left|\mathbf{l}_{2}\right|$,
and $\left|\mathbf{l}_{12}\right|$, each of them belongs to a distinct
$\mathbb{R}_{i}^{1}$, formed a closed triangle $\mathbf{a}$, which
is in fact a \textit{portion} of a flat 2-dimensional space $\mathbb{R}^{2}.$
Then we can think segment $\left|\mathbf{l}_{1}\right|$, $\left|\mathbf{l}_{2}\right|$,
and $\left|\mathbf{l}_{12}\right|$ are embedded on a common $\mathbb{R}^{2}.$
Using this triangle, we can construct 2-dimensional vectors describing
each segments: $\mathbf{l}_{1}$, $\mathbf{l}_{2}$, and $\mathbf{l}_{12}$
$\in T_{p}M\sim\mathbb{R}^{2}.$ In other words, we assume the segments
$\mathbf{l}_{1}$, $\mathbf{l}_{2}$, and $\mathbf{l}_{12}$ are embedded
in a one dimensional higher, flat manifold, which is the triangle
$\mathbf{a}$. This assumption is reasonable because before doing
coarse-graining on the slice, firstly we need to coarse-grain the
bulk, which means the bulk where the fine-grained slice lied must
already be flatten \cite{key-4.3}. Using the Minkowski theorem in
$\mathbb{R}^{2}$, we have $\mathbf{l}_{1}+\mathbf{l}_{2}+\mathbf{l}_{12}=0$,
where $\mathbf{l}_{1}$, $\mathbf{l}_{2}$, and $\mathbf{l}_{12}$
formed a closed triangle.}
\end{figure}
 Doing the same way to any microstate $\left\{ \left|\mathbf{l}_{+}^{\left(j\right)}\right|,\left|\mathbf{l}_{-}^{\left(j\right)}\right|,\alpha^{\left(j\right)}\right\} ,$
we obtain all possible \textit{reduced} \textit{$j^{th}$-microstate},
$\left|\mathbf{l}_{+}^{\left(j\right)}\right|$:
\begin{equation}
\left|\mathbf{l}_{+}^{\left(j\right)}\right|=\sqrt{\left|\mathbf{l}_{1}^{\left(j\right)}\right|^{2}+\left|\mathbf{l}_{2}^{\left(j\right)}\right|^{2}-2\left|\mathbf{l}_{1}^{\left(j\right)}\right|\left|\mathbf{l}_{2}^{\left(j\right)}\right|\cos\phi_{12}^{\left(j\right)}}.\label{eq:4.4}
\end{equation}
 We define the \textit{macrostate} $\overline{\left|\mathbf{L}\right|}$
as an \textit{average value} \textit{over all possible configuration
under a weight/probability distribution $p_{j}:$}
\begin{equation}
\overline{\left|\mathbf{L}\right|}=\sum_{j}p_{j}\left|\mathbf{l}_{+}^{\left(j\right)}\right|,\qquad j\in\mathbb{Z}.\label{eq:macro}
\end{equation}
The \textit{coarse-grained length} is defined as the macrostate, relative
to \textit{fine-grained length}, which is defined to be \textit{sum
of the norms} of the $\mathbf{l}_{1}$ and $\mathbf{l}_{2}$:
\begin{equation}
\left|\mathbf{l}\right|_{+}=\left|\mathbf{l}_{1}\right|+\left|\mathbf{l}_{2}\right|.\label{eq:4.4a}
\end{equation}
See FIG. 4. 

In the statistical mechanics point of view, we can only have certain
information about the macrostate $\overline{\left|\mathbf{L}\right|}$,
without knowing the details of the fine degrees of freedom, i.e.,
we do not know the correct values of $\left\{ \left|\mathbf{l}_{1}\right|,\left|\mathbf{l}_{2}\right|,\phi_{12}\right\} $,
in this case. But we can assume them to be some numbers, so that they
give the correct average value under a specific probability distribution,
which is the macrostate $\overline{\left|\mathbf{L}\right|}$. 

In the end, by taking the full-discretized object mentioned in the
beginning of this section, we obtain the coarse-grained spatial slice;
$\mathcal{S}_{\triangle=3}^{1}$, which is the macrostate: an average
over all possible fine-grained slices, the microstates $\mathcal{S}_{\triangle=6}^{1}$.

\subsection{Restrictions}

\subsubsection{Bounds on informational entropy}

We have already reviewed the definition of informational entropy in
Subsection II B. Now we will apply this to coarse-grained discrete
geometries. We only use Shannon entropy, since we are dealing with
discrete system.

\paragraph{Lower bound entropy.}

To prevent divergencies in the infinitely-small scale limit, as explained
in Subsection II B, we need to give \textit{first restriction} to
the sample space $\mathbb{X}_{\mathbf{L}}.$ This results in the countability
of all the possible microstates, as well as the discreteness in the
sample space. The physical consequence of our first restriction is
that all the physical properties -in this case, the properties of
the microstates $\left\{ \left|\mathbf{l}_{1}\right|,\left|\mathbf{l}_{2}\right|,\phi_{12}\right\} ,$
must be discrete. This statement is equivalent with the statement:
$\left\{ \left|\mathbf{l}_{1}\right|,\left|\mathbf{l}_{2}\right|,\phi_{12}\right\} $
can not be infinitesimally-small, there should exist an \textit{ultraviolet
cut-off} which prevent the lengths and angles to be arbitrarily small.
This is not a problem in Regge geometry since all the geometrical
objects are defined by simplices, which are discrete objects. 

We consider the \textbf{Kronecker-delta} distribution as the probability
distribution describing a completely-known system with maximal certainty,
-a '\textit{pure} state'. The Shannon entropy of this pure state is
zero, by (\ref{eq:3.0a}), which is consistent with its physical interpretation. 

Therefore, the lower bound of the informational entropy is zero, resulting
from the Kronecker-delta probability distribution. This is in analog
with the von-Neumann entropy bound on the pure state for the spin-network
calculation studied in \cite{key-4.3}.

\paragraph{Upper-bound entropy. }

To prevent divergencies in the infinitely-large scale limit, we need
to give \textit{second restriction} to the sample space $\mathbb{X}_{\mathbf{L}}.$
Following the reasoning described in Subsection II B, giving the second
restriction to our discrete geometry system will results in the finiteness
of the number of possible microstates, besides of being countable,
due to the first restriction. The physical implication of the second
restriction, along with the first restriction, is the properties of
microstates $\left\{ \left|\mathbf{l}_{1}\right|,\left|\mathbf{l}_{2}\right|,\phi_{12}\right\} ,$
can \textit{not} be infinitely large, nor infinitely small. This agrees
with the atomism principle: the 'atoms' of space can not be infinitely
large. 

We consider the homogen distribution (\ref{eq:3.1}), as the probability
distribution describing a completely-unknown system with maximal uncertainty,
-a \textit{'maximally-mixed'} state. With the first and second restrictions,
the entropy of the homogeneous probability distribution from (\ref{eq:9})
which is $S_{\textrm{max}}=\ln W,$ could be obtained.

In conclusion, with first and second restrictions, we have a well-defined
entropy for any probability distribution for our discrete geometry
system, within the range defined in (\ref{eq:y}). See Section II
B for a detailed derivation concerning bounds on entropy.

\subsubsection{The cut-offs and truncation to the theory}

Both the first and second restriction define cut-offs to the properties
of the system. To give a well-defined and a consistent lower and upper
bound of entropy for the maximal and minimal certainty cases, the
sample space $\mathbb{X}_{\mathbf{L}}$ needs to be discrete and finite.
Because of this reason, the properties of microstates $\left\{ \left|\mathbf{l}_{1}\right|,\left|\mathbf{l}_{2}\right|,\phi_{12}\right\} ,$
can not be infinitely large, nor infinitely small. These can be geometrically
interpreted as a restriction to the range of the length of the segments:
the segments can not be infinitely small (they must be discrete) and
can not be infinitely large: 
\begin{eqnarray}
|\mathbf{l}_{\textrm{min}}|\leq|\mathbf{l}|\leq|\mathbf{l}_{\textrm{max}}|, & \qquad & \mathbf{l}\in\Sigma;\; f:\Sigma\rightarrow\mathbb{Z},\label{eq:discrete}\\
|\mathbf{l}_{\textrm{min}}|>0, & \qquad & |\mathbf{l}_{\textrm{max}}|<\infty.\nonumber 
\end{eqnarray}
$f$ is a bijective map which maps every elements of $\Sigma$ to
the set of integer numbers $\mathbb{Z}$. The minimal and maximal
values $|\mathbf{l}_{\textrm{min}}|$ and $|\mathbf{l}_{\textrm{max}}|$
are the \textit{ultraviolet} and the\textit{ infrared cut-offs} which
prevent the theory from divergencies. Setting these values, we define
a \textit{truncation} to our theory. Returning back to the discretization
of $\mathcal{S}^{1}$ in previous subsection, with these truncations,
the refinement must stop at some point; it can not go to arbitrarily
small details, and the 'perfect' circle $\mathcal{S}^{1}$ can never
be reached, since it is only an idealization.

\subsection{Microcanonical case}

As explained in Section II C, the \textit{microcanonical ensemble}
is an ensemble of system which satisfies constraints $\Delta E=0,$
$\triangle N=0$; the amount of energy and number of particles in
each microstates are fixed. In analog to the kinetic theory, the reduced-microstates
(\ref{eq:4.4}) must satisfies the \textit{microcanonical constraint}
for a system of coupled-segments: all the possible reduced-microstate
must be equal to the correct reduced-microstate: 
\begin{equation}
\left|\mathbf{l}_{+}^{\left(j\right)}\right|=\left|\mathbf{l}_{12}\right|.\label{eq:4-6}
\end{equation}
It must be noted that with this constraint, the values of $\left|\mathbf{l}_{1}^{\left(j\right)}\right|$,
$\left|\mathbf{l}_{2}^{\left(j\right)}\right|$, $\phi_{12}^{\left(j\right)}$
could vary, but their relation (their 'total length') described in
(\ref{eq:4.4}) must be equal to the correct reduced microstate $\left|\mathbf{l}_{12}\right|.$
This also can be geometrically interpreted as taking the 'total length'
and number of partition (which is two fine-grained segments for each
coarse-grained segment) to be constants.

To conclude this subsection, coarse-graining from $\mathcal{S}_{\triangle=6}^{1}$
to $\mathcal{S}_{\triangle=3}^{1}$ (FIG. 3) means we only \textit{know}
with certainty the macrostate $\overline{\left|\mathbf{L}\right|}$,
and lose the information about the correct configuration of microstate
$\left\{ \left|\mathbf{l}_{1}^{\left(j\right)}\right|,\left|\mathbf{l}_{2}^{\left(j\right)}\right|,\right.$
$\left.\phi_{12}^{\left(j\right)}\right\} $ (which is $\left\{ \left|\mathbf{l}_{1}\right|,\left|\mathbf{l}_{2}\right|,\phi_{12}\right\} ,$
respectively). See FIG. 5. 
\begin{figure}[h]
\centerline{\includegraphics[width=5.8cm]{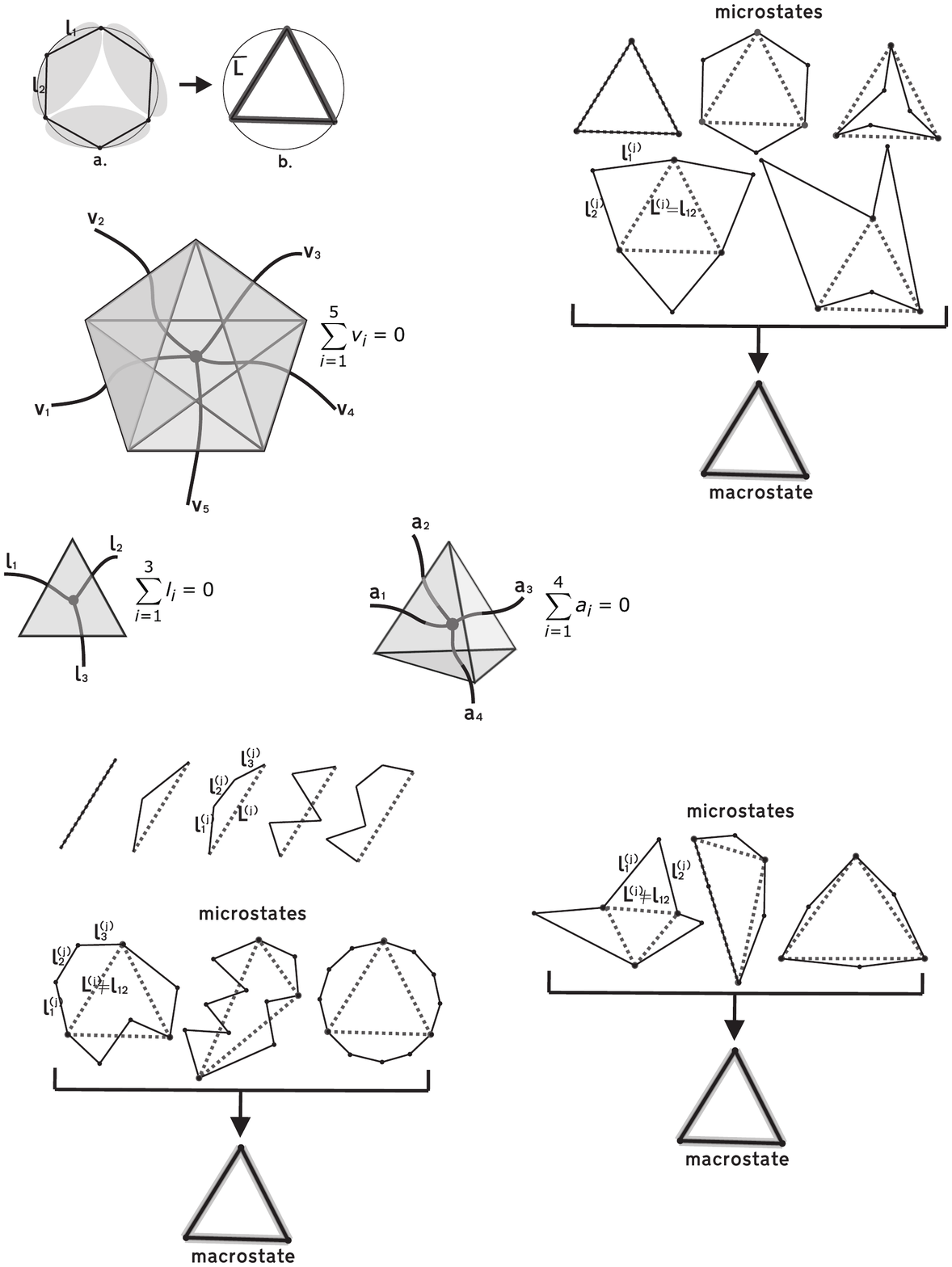}} \caption{The microcanonical case: Some of the possible microstates $\mathcal{S}_{\triangle=6}^{1}$
which contains six partitions, coarse-grained into a single macrostate
$\mathcal{S}_{\triangle=3}^{1}$ , which contains only three partitions.
The total length of each two adjacent segments $\left\{ \mathbf{l}_{1}^{\left(j\right)},\mathbf{l}_{2}^{\left(j\right)}\right\} $
are equal for each microstates, which is $\mathbf{l}_{12}$.}
\end{figure}

\subsection{General ensemble case}

\subsubsection{Canonical case and length fluctuation}

The \textit{canonical ensemble} in kinetic theory is an ensemble of
systems which only satisfy the constraint $\Delta N=0.$ The energy
can flow in and out from each microstate, but the number of partitions
is conserved. Using the same analogy here, we define in general the
canonical ensemble in the discrete geometrical sense by relaxing the
microcanonical constraint (\ref{eq:4-6}); in the canonical case,
$\left|\mathbf{l}_{+}^{\left(j\right)}\right|$ does not need to be
equal to $\left|\mathbf{l}_{12}\right|$: 
\begin{equation}
\left|\mathbf{l}_{+}^{\left(j\right)}\right|=\left|\sum_{i=1}^{2}\mathbf{l}_{i}^{\left(j\right)}\right|.\label{eq:13}
\end{equation}
This means the total length of two adjacent segments may vary, as
long as averaging all of them still give the macrostate $\overline{\left|\mathbf{L}\right|},$
see FIG. 6.

\begin{figure}[h]
\centerline{\includegraphics[width=6cm]{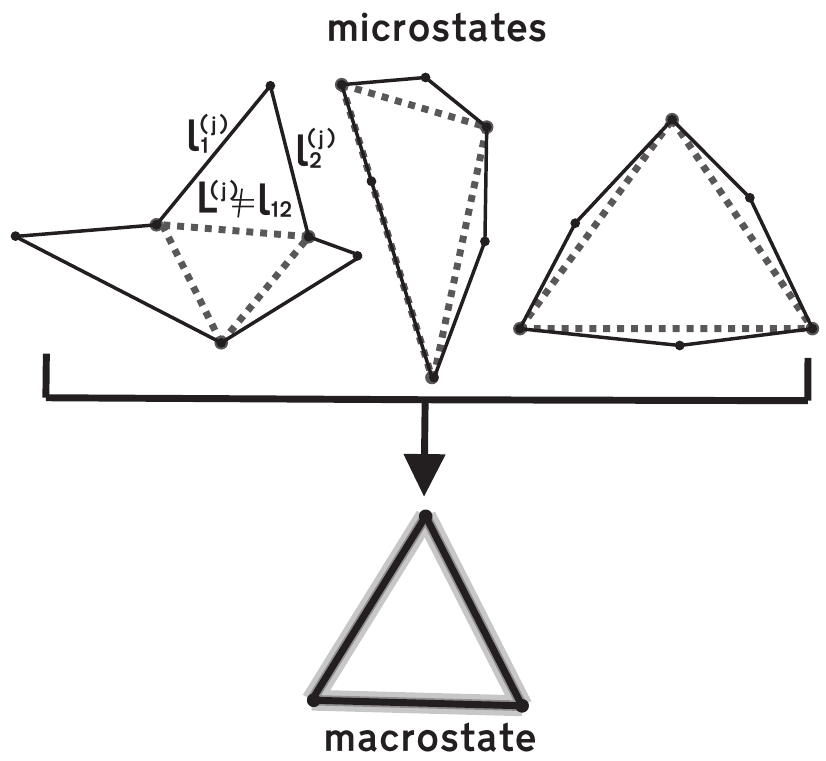}} \caption{The canonical case: Some of the possible microstates $\mathcal{S}_{\triangle=6}^{1}$
which contains six partitions, coarse-grained into a single macrostate
$\mathcal{S}_{\triangle=3}^{1}$ , which contains only three partitions.
The total length of each two adjacent segments $\left\{ \mathbf{l}_{1}^{\left(j\right)},\mathbf{l}_{2}^{\left(j\right)}\right\} $
may vary. This will lead to a fluctuation of length, since now the
microstates $\left|\mathbf{l}_{+}^{\left(j\right)}\right|_{c}$ is
not a constant.}
\end{figure}

Without the microcanonical constraint, we have more possible microstates,
while the first and second restriction are still maintained. In this
case, we have fluctuations of length. The variance of the length,
in general, is not zero; this will depend on the probability density.
It is clear that for the Kronecker delta, the variance is zero, no
matter if the case is canonical or microcanonical. This condition
is consistent with the physical interpretation, since in the Kronecker
delta distribution, we only have one possible microstate, so there
is no fluctuation of length.

\subsubsection{Grandcanonical case, length, and 'particle' fluctuations}

The most general ensemble in kinetic theory is the \textit{grandcanonical
ensemble}, where we have no restriction on $\Delta E$ and $\Delta N$.
In the two previous ensembles, the number of partitions is fixed:
two partitions for each coarse-grained segment of the discretization.
Now, we release this restriction; the relation (\ref{eq:13}) of the
canonical case becomes: 
\begin{equation}
\left|\mathbf{l}_{+}^{\left(j,n_{j}\right)}\right|_{g}=\left|\sum_{i=1}^{n_{j}}\mathbf{l}_{i}^{\left(j\right)}\right|,\label{eq:14}
\end{equation}
which means, each different microstate $\left|\mathbf{l}_{+}^{\left(j\right)}\right|_{g}$
will have degeneracy in terms of number of partitions. See FIG. 7.

\begin{figure}[h]
\centerline{\includegraphics[scale=0.85]{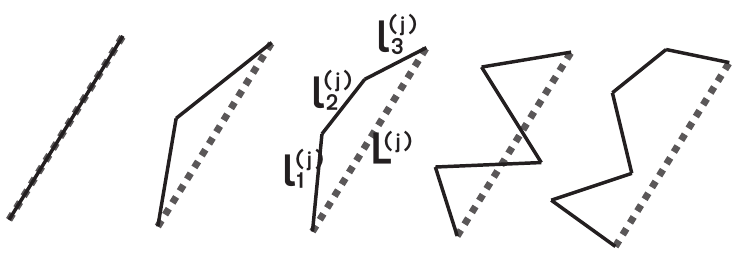}} \caption{Each different microstate $\left|\mathbf{l}_{+}^{\left(j\right)}\right|_{g}$
has degeneracy in terms of number of partitions, which is the number
of segments.}
\end{figure}

The microstate of the grandcanonical system also depends on the number
of partitions $n_{j}$, which in statistical mechanics, is usually
called as the \textit{occupation number}. The average length is given
by: 
\[
\overline{\left|\mathbf{L}\right|}=\sum_{j}^{W}P_{j}\left|\mathbf{l}_{+}^{\left(j,n_{j}\right)}\right|_{g},
\]
with $P_{j}$ is the grand-canonical distribution function, $W$ is
the number of all possible microstates, and $n_{j}$ is the number
of partition on each microstate $\left|\mathbf{l}_{+}^{\left(j\right)}\right|_{g}.$
We can collect all microstates having the same occupation numbers
$n_{j}$. The idea is similar to collecting states which have same
wave-number $k$ when we define the Fock space in quantum mechanics.
Furthermore, we could defined the average occupation number: 
\[
\bar{n}=\sum_{k}^{\mathcal{W}}\mathcal{P}_{k}n_{k}.
\]
The probability distribution $\mathcal{P}_{k}$ is 'Fourier conjugate'
to $P_{j}$. 

Obviously, in the grand canonical case, we have length fluctuation,
using the same definition as in the canonical case. But there is also
another fluctuation: the fluctuation of the number of partitions,
or, in the statistical mechanics language, 'particle number' fluctuation.
The variance, standard deviation, and fluctuation can be obtained
using the standard formula; if $P_{j}$ is the Kronecker-delta distribution,
then the variance this is similar to the previous cases. See FIG.
8. 
\begin{figure}[h]
\centerline{\includegraphics[width=6cm]{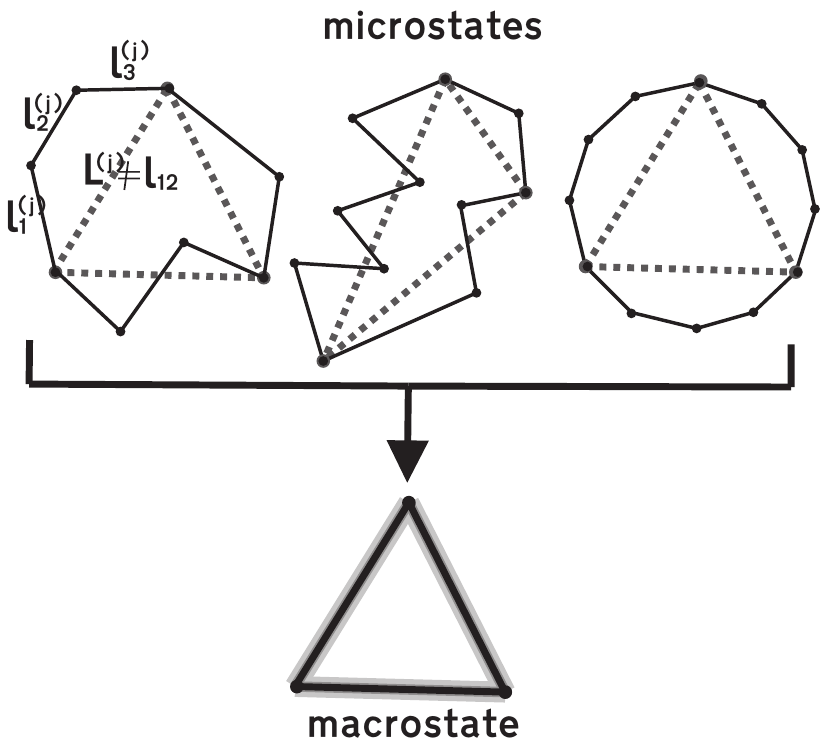}} \caption{The grandcanonical case: Some of the possible microstates $\mathcal{S}_{\triangle=6}^{1}$
which can contain any number of partitions, coarse-grained into a
single macrostate $\mathcal{S}_{\triangle=3}^{1}$, which contains
only three partitions. This will lead to a fluctuation in the number
of partitions, since now the number of partition of each microstates
is not a constant. The fluctuation of length also exist, in general.}
\end{figure}

\subsection{Higher dimensional case}

In the previous sections, we worked in the (1+1)-dimensional case,
and the geometrical object we coarse-grained were collections of segments,
a 1-dimensional slice of a 2-dimensional spacetime. In higher dimensions,
we could coarse-grain areas and volumes.

In this subsection, we will use the construction of \textit{$n$-simplex}
procedure explained in our earlier work \cite{key-7} as a basic property.
Another important theorem we will used in this subsection is the \textit{Minkowski
theorem} or the closure constraint \cite{key-3.5,key-3.26}. It is
related to the construction of $p$-simplices embedded in $\mathbb{R}^{n}$.
For $p=2,$ it is a construction of a closed triangle from three vectors
$\left\{ \mathbf{l}_{1},\right.$ $\mathbf{l}_{2},$ $\left.\mathbf{l}_{3}=-\mathbf{l}_{12}\right\} \in\mathbb{R}^{n},$
which satisfy the closure constraint: 
\[
\sum_{i=1}^{3}\mathbf{l}_{i}=0.
\]
Using physical terminology, this constraint is called the \textit{gauge
invariance condition}. It is already explained in \cite{key-7} that
closure condition will cause triangle-inequality, but not vice-versa.
See FIG. 4  as an example. We will use this property to generalize
our result in higher dimension.

\subsubsection{(2+1) D: Coarse-graining area}

In the (2+1) D theory, the slice $\Sigma$ of a 3D manifold $\mathcal{M}$
is a 2-dimensional surface, therefore in this case, we will be coarse-graining
the area of surfaces. Let a discretized surface be triangulated by
three triangles as illustrated in FIG. 9(a). 
\begin{figure}[h]
\centerline{\includegraphics[scale=0.85]{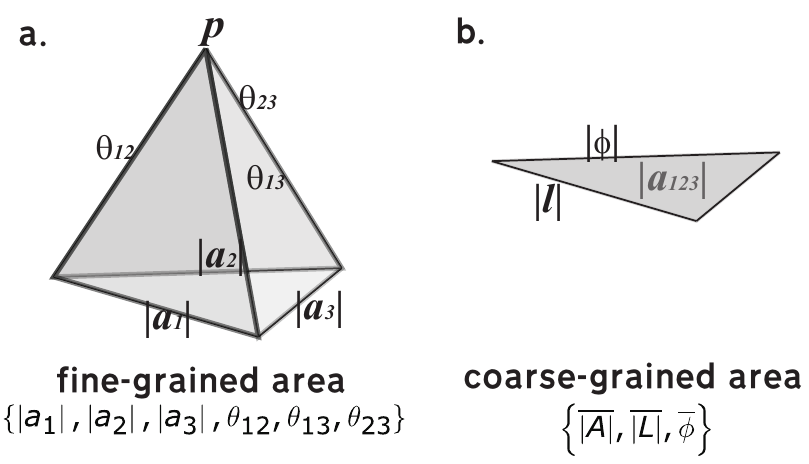}} \caption{(a) is a portion of a discretized surface triangulated by three triangles.
Coarse-graining the area of the surface is defined as treating these
three triangles as one system: by substituting them with a single
and flat triangle in figure (b), while keeping the boundary (segments
$\left|\mathbf{l}_{1}\right|$, $\left|\mathbf{l}_{2}\right|$, $\left|\mathbf{l}_{3}\right|$)
fixed. This transformation is known as 3-1 Pachner move.}
\end{figure}
 Let us apply the 3-1 Pachner move. This configuration contains six
degrees of freedom, we can choose them to be written as $\left\{ \left|\mathbf{a}_{1}\right|,\left|\mathbf{a}_{2}\right|,\left|\mathbf{a}_{3}\right|,\theta_{12},\theta_{23},\theta_{13}\right\} $
to prevent ambiguities: the system contains three triangles with area
$\left|\mathbf{a}_{i}\right|$ and coupling constant $\theta_{ij}$
between each two of them \cite{key-7}. From the information of $\left\{ \theta_{12},\theta_{23},\theta_{13}\right\} $
(the angles between two triangles, -\textit{3D dihedral angles}),
using the inverse dihedral angle relation \cite{key-7}:
\begin{equation}
\cos\phi_{ij}=\frac{\cos\theta_{ij,k}-\cos\theta_{ik,j}\cos\theta_{kj,i}}{\sin\theta_{ik,j}\sin\theta_{kj,i}},\label{eq:inverse}
\end{equation}
we could obtain $\left\{ \phi_{12},\phi_{23},\phi_{13}\right\} $
(the angles between two segments, -\textit{2D dihedral angles}), and
then calculate the deficit angle on the hinge (the point $p$ in FIG.
9(a)): 
\[
\delta\phi=2\pi-\sum_{i=1}^{3}\phi_{i},
\]
which defines the intrinsic 2D curvature of the surface. These information
\textit{completely} determine the discretized surface. It must be
noted that we did \textit{not} use vectorial objects to define the
surface, all the variables we have are written in a coordinate-free
way, i.e., they are scalars. See FIG. 10. 
\begin{figure}[h]
\centerline{\includegraphics{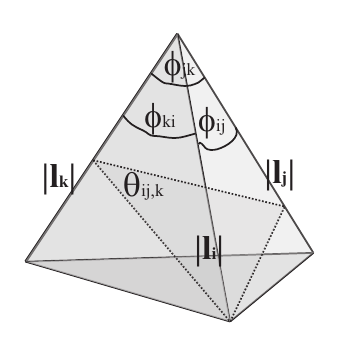}} \caption{Given angles $\phi_{ij}$, $\phi_{ik}$, $\phi_{jk}$ at point $p$
of a tetrahedron, we could obtain the dihedral angle $\theta{}_{ij,k}$.
In fact, $\theta{}_{ij,k}$ is only $\phi_{ij}$ projected on the
plane normal to segment $\left|\mathbf{l}_{k}\right|$.}
\end{figure}

The next step is to obtain 'transformation' to the 'center-of-mass'
variables, the 'center-of-mass' of the system is a single triangle,
obtained by 'summing-up' the three triangles of 3-1 Pachner move.
This 'center-of-mass' variables can be written as $\left\{ \left|\mathbf{a}_{123}\right|,\left|\mathbf{l}\right|,\phi,\boldsymbol{\alpha}\right\} ,$
with $\left|\mathbf{a}_{123}\right|$ is the 'total area': the \textit{norm
of the sum} of areas $\mathbf{a}_{1}$, $\mathbf{a}_{2}$, and $\mathbf{a}_{3}$;
$\left|\mathbf{l}\right|$ and $\phi$ are, respectively, one of the
segment and 2D angle of this 'center-of-mass' triangle, and $\boldsymbol{\alpha}$
are the variables which we are going to coarse-grained (we could choose
it to be $\boldsymbol{\alpha}=\left(\left|\mathbf{a}_{12}\right|,\left|\mathbf{a}_{1}\right|,\alpha\right),$
with $\alpha$ is the angle between triangles $\left|\mathbf{a}_{12}\right|$
and $\left|\mathbf{a}_{1}\right|$, but they are not so important
since we are going to neglect them all by the coarse-graining procedure). 

For the next step, we would like to obtain the formula relating the
total area $\left|\mathbf{a}_{123}\right|$ with the fine-grained
areas $\left|\mathbf{a}_{i}\right|.$ Let us return to the vectorial
notation $\mathbf{a}_{i}$. In the vectorial picture, it is obvious
that the three triangles of the discretized surface \textit{and} the
triangle of the total area can be arranged so that they construct
a flat tetrahedron, in the same way four 2-simplices constructed a
3-simplex. This flat tetrahedron is embedded in $\mathbb{R}^{3}$
(in fact, it \textit{is} a portion of $\mathbb{R}^{3}$). The three
triangles and the total triangle form the surface of the tetrahedron
(which is automatically embedded on $\mathbb{R}^{3}$). Therefore,
we could describe these triangles vectorially, using a 2-form $\mathbf{a}_{i}\in\Omega^{2}\left(\mathbb{R}^{3}\right)$
(with $\Omega^{2}\left(\mathbb{R}^{3}\right)$ is the space of $2$-form
over $\mathbb{R}^{3}$), such that their norms give the same norms
$\left|\mathbf{a}_{i}\right|$ as before. Using the closure constraint
(or Minkowski theorem in $\mathbb{R}^{3}$ ($\Omega^{2}\left(\mathbb{R}^{3}\right)$
is isomorphic to $\mathbb{R}^{3}$)), we have the gauge invariance
condition, for the four triangles to form a closed tetrahedron in
$\mathbb{R}^{3}$: 
\begin{equation}
\sum_{i=1}^{4}\mathbf{a}_{i}=0,\label{eq:19}
\end{equation}
which in our case, can be written as: 
\begin{equation}
\mathbf{a}_{4}=-\mathbf{a}_{123}=-\left(\mathbf{a}_{1}+\mathbf{a}_{2}+\mathbf{a}_{3}\right).\label{eq:15}
\end{equation}
See FIG. 11. 
\begin{figure}[h]
\centerline{\includegraphics[scale=0.8]{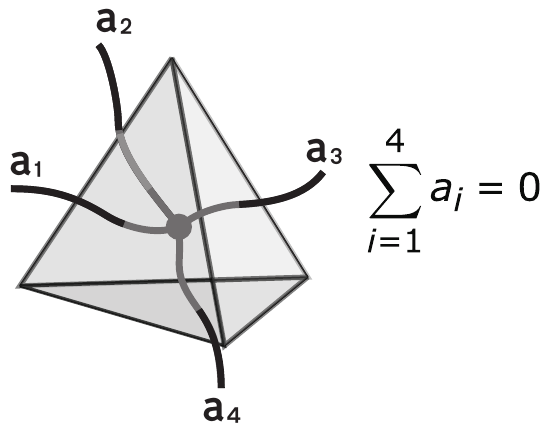}} \caption{The Minkowski theorem in $\mathbb{R}^{3}$, a closed surface of a
tetrahedron must satisfy gauge invariance condition.}
\end{figure}

Taking the norm of equation (\ref{eq:15}), the 'transformation' relation
we would like to obtain is:
\begin{eqnarray}
\left|\mathbf{a}_{123}\right| & = & \sqrt{\sum_{i=1}^{3}\left|\mathbf{a}_{i}\right|^{2}-2\sum_{\binom{i<j}{i=1}}^{3}\left|\mathbf{a}_{i}\right|\left|\mathbf{a}_{j}\right|\cos\theta_{ij}},\label{eq:16}
\end{eqnarray}
\[
\boldsymbol{\alpha}=\boldsymbol{\alpha}\left(\left|\mathbf{a}_{1}\right|,\left|\mathbf{a}_{2}\right|,\left|\mathbf{a}_{3}\right|,\theta_{12},\theta_{23},\theta_{13}\right),
\]
with $\theta_{ij}$ are the 3D dihedral angles. The other variables
$\left|\mathbf{l}\right|,\phi$ can be obtained given the information
of $\left\{ \left|\mathbf{a}_{1}\right|\right.,$ $\left|\mathbf{a}_{2}\right|,$
$\left|\mathbf{a}_{3}\right|,$ $\theta_{12}$, $\theta_{23},$ $\left.\theta_{13}\right\} $,
see \cite{key-7}. Similar with the lower dimension analog, we use
the vectorial form just as a simple way to derive this transformation.

At a first glance, formula $\left(\ref{eq:16}\right)$ seems to be
an 'extrinsic' relation, since $\cos\theta_{ij}$ is an extrinsic
property, relative to the 2D surface (the angle $\theta_{ij}$ does
not 'belong' to the 2D surface $\Sigma$). But by using the remarkable
\textit{dihedral angle formula}, the 3D dihedral angles of the tetrahedron
can be written as functions of the angles between segments using its
inverse formula (\ref{eq:inverse}). See FIG. 10. For a detailed explanation
about these angles, see \cite{key-7,key-3.28,key-3.29}. Therefore,
it is clear that formula $\left(\ref{eq:16}\right)$ is \textit{purely
intrinsic} to the 2D surface, it does \textit{not} depend on the embedding
in $\mathbb{R}^{3},$ we only use $\mathbb{R}^{3}$ to help us to
derive relation (\ref{eq:16}) in an easy way.

Now, as a 2-dimensional analog to the coarse-graining in 1-dimension
carried in Subsection III A, we define the statistical mechanics terminologies.
Let us called $\left\{ \left|\mathbf{a}_{123}\right|,\left|\mathbf{l}\right|,\phi,\boldsymbol{\alpha}\right\} $
as the \textit{correct microstate}, and $\left\{ \left|\mathbf{a}_{123}^{\left(k\right)}\right|,\left|\mathbf{l}^{\left(k\right)}\right|,\phi^{\left(k\right)},\boldsymbol{\alpha}^{\left(k\right)}\right\} ,$
as the \textit{$k^{th}$-microstate}. We reduce the degrees of freedom
away by removing $\boldsymbol{\alpha}$ and defining the \textit{correct
reduced-microstate} as $\left\{ \left|\mathbf{a}_{123}\right|,\left|\mathbf{l}\right|,\phi\right\} ,$
and the $k^{th}$ \textit{reduced-microstate} as $\left\{ \left|\mathbf{a}_{123}^{\left(k\right)}\right|,\left|\mathbf{l}^{\left(k\right)}\right|,\phi^{\left(k\right)}\right\} ,$
satisfying:

\begin{equation}
\left|\mathbf{a}_{123}^{\left(k\right)}\right|=\sqrt{\sum_{i=1}^{3}\left|\mathbf{a}_{i}^{\left(k\right)}\right|^{2}-2\sum_{\binom{i<j}{i=1}}^{3}\left|\mathbf{a}_{i}^{\left(k\right)}\right|\left|\mathbf{a}_{j}^{\left(k\right)}\right|\cos\theta_{ij}^{\left(k\right)}},\label{eq:17}
\end{equation}
\begin{eqnarray*}
\left|\mathbf{l}^{\left(k\right)}\right| & = & \left|\mathbf{l}_{12}^{\left(k\right)}\right|,\\
\phi^{\left(k\right)} & = & f\left(\left|\mathbf{l}_{12}^{\left(k\right)}\right|,\boldsymbol{\alpha}^{\left(k\right)}\right).
\end{eqnarray*}
The \textit{coarse-grained triangle} is defined as the \textit{macrostate}
of the system $\left\{ \overline{\left|\mathbf{A}\right|},\overline{\left|\mathbf{L}\right|},\bar{\phi}\right\} $
: 
\begin{eqnarray*}
\overline{\left|\mathbf{A}\right|} & = & \sum_{k}p_{k}\left|\mathbf{a}_{123}^{\left(k\right)}\right|,\\
\overline{\left|\mathbf{L}\right|} & = & \sum_{k}p_{k}\left|\mathbf{l}^{\left(k\right)}\right|,\\
\bar{\phi} & = & \sum_{k}p_{k}\phi^{\left(k\right)},
\end{eqnarray*}
which is an average value over all possible reduced-microstates (\ref{eq:17}).
We called $\overline{\left|\mathbf{A}\right|}$ as the \textit{coarse-grained
area}, while the\textit{ fine-grained area} is the 'sum of the norms':
$\sum_{i=1}^{3}\left|\mathbf{a}_{i}\right|$; the area of the 'correct'
microstate. The microcanonical case is described by the constraint:
\[
\left|\mathbf{a}_{123}^{\left(k\right)}\right|=\left|\mathbf{a}_{123}\right|.
\]

Now, what is the geometrical interpretation of this coarse-graining
method? First, we look at formula (\ref{eq:16}), which looks like
the total 'lagrangian' of the system: $\sum_{i=1}^{3}\left|\mathbf{a}_{i}\right|^{2}$
looks like the 'kinetic' part, while the $\left|\mathbf{a}_{i}\right|\left|\mathbf{a}_{j}\right|\cos\theta_{ij}$
looks like the 'interaction' part between fields $\left|\mathbf{a}_{i}\right|$
and $\left|\mathbf{a}_{j}\right|$, with $\cos\theta_{ij}$ as the
'coupling constant'. We can think the discretized surface as a many-body
system containing (in this case) three 'quanta' of area, with interactions
among them. The measure of how strong are the interactions is described
by the 'couplings': the angles between these quanta of area, which
in turn, by (\ref{eq:inverse}), is related to the intrinsic curvature
of the surface. In this sense, we could think the intrinsic curvature
as an \textit{emergent} property of a many-body system; \textit{which
measures the intensity of the interaction among the 'quanta' of areas
of the surface} \cite{key-7}. Small curvature (small deficit angle,
which means the coupling, -the cosine of the dihedral angle between
two 'quanta' $\approx1$) gives maximal interaction between the 'quanta'
of space, while large curvature (large deficit angle, which means
the coupling $\approx0$) gives minimal interaction.

Second, by coarse-graining, we treat the system (in our case) as a
single 'particle' of a area, which means we loose the information
about the individual ``atoms'' and also the interaction among them.
This interpretation is consistent with the fact that by coarse-graining,
we loose the information about the intrinsic curvature of the slice,
since a collection of several 'interacting' triangles is replaced
by a single flat triangle. See FIG. 12  for a coarse-grained area
example. 

\begin{figure}[h]
\centerline{\includegraphics[width=6cm]{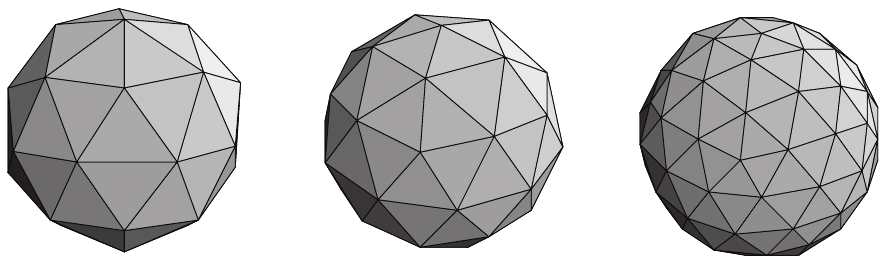}} \caption{Coarse-graining areas will correspond in loosing the intrinsic curvature
of the surface.}
\end{figure}

As a last point, we need to generalize these results to a more general
statistical ensembles: for the canonical case we have $\left|\mathbf{a}_{123}^{\left(k\right)}\right|=\left|\sum_{i=1}^{3}\mathbf{a}_{i}^{k}\right|,$
while for the grandcanonical case we have $\left|\mathbf{a}_{123}^{\left(k\right)}\right|=\left|\sum_{i=1}^{n_{k}}\mathbf{a}_{i}^{k}\right|$,
both are, in general, not equal to $\left|\mathbf{a}_{123}\right|$.
The area and particle number fluctuations are defined in the same
way as in the 1-dimensional case. We should remember that we still
have freedom to coarse-grain length just as in the previous section,
but it can only be done \textit{after} we coarse-grain the surface.

\subsubsection{(3+1) D: Coarse-graining volume}

Let us go to the real world by adding one dimensional higher. In (3+1)
dimensions we will discuss two cases: the 4-1 and 3-2 moves. The slice
$\Sigma$ of a 4D manifold $\mathcal{M}$ is a 3-dimensional space,
therefore in this case, we will coarse-grain volumes of space.

\paragraph{4-1 Pachner moves case.}

We take a portion of 3D space discretized by four tetrahedra, illustrated
in FIG. 13, which is the 4-1 Pachner move.

\begin{figure}[h]
\centerline{\includegraphics[scale=0.85]{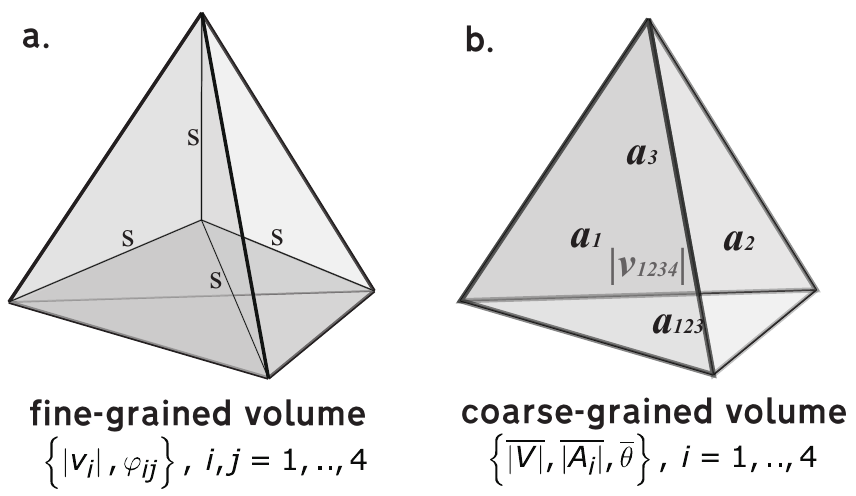}} \caption{(a) is a portion of a discretized space triangulated by four tetrahedra.
Coarse-graining the volume of the space is defined as treating these
four tetrahedra as one system: by substituting them with a single,
flat tetrahedron in figure (b) while keeping the boundary (triangles
$\mathbf{a}_{1}$, $\mathbf{a}_{2}$, $\mathbf{a}_{3}$, $\mathbf{a}_{4}$)
fixed. This transformation is known as 4-1 Pachner move.}
\end{figure}

Each tetrahedron is a portion of an $\mathbb{R}^{3}$ space, which
is distinct for each tetrahedron. In the vectorial way, they are described
by 3-forms $\mathbf{v}_{i}$, obtained from the wedge product of three
segments $\mathbf{l}_{j}$ (see \cite{key-7}), or it can be obtained
from the area of the triangles $\mathbf{a}_{j}$. The volume of the
tetrahedron is defined by its norm, or by: 
\begin{equation}
\left|\mathbf{v}_{i}\right|^{2}=\frac{2}{9}\mathbf{a}_{j}\cdot\left(\mathbf{a}_{k}\times\mathbf{a}_{l}\right),\label{eq:20}
\end{equation}
with $\mathbf{a}_{j}\in\Omega^{2}\left(\mathbb{R}^{3}\right)$ are
the 2-simplices: three triangles from all the four which bound the
tetrahedron (the closure constraint on $\mathbb{R}^{3}$ (\ref{eq:19})
guarantees any combination will give the same volume). It must be
emphasized here that we are working on a purely Regge geometry picture
and we are not going to consider the \textit{twisted geometry} case
in this work. This means we know \textit{exactly} the shape of the
tetrahedra; not only the norm of the areas $\left|\mathbf{a}_{j}\right|$
of the triangles (as in the twisted geometry case), but \textit{also}
the length of each segments of the tetrahedron $\left|\mathbf{l}_{j}\right|$.
This is equivalent with having the information about the vectorial
triangles $\mathbf{a}_{j}$, which are used to derive the volume of
the tetrahedron.

The system in FIG. 13(a) is described by ten degrees of freedom (see
\cite{key-7}), we choose the variables to be the individual volumes
of tetrahedra and the 4D dihedral angles between them $\left\{ \left|\mathbf{v}_{i}\right|,\varphi_{ij}\right\} $,
for $i,j=1,2,3,4,\, i\neq j$. Similar to the previous lower-dimensional
cases, the next step is to obtain the center-of-mass variables, which
are $\left\{ \left|\mathbf{v}_{1234}\right|,\left|\mathbf{a}_{i}\right|,\theta,\boldsymbol{\alpha}\right\} ,$
for $i=1,..,4$. $\left\{ \left|\mathbf{v}_{1234}\right|,\left|\mathbf{a}_{i}\right|,\theta\right\} $
describe the center-of-mass tetrahedron: $\left|\mathbf{v}_{1234}\right|$
is the volume, $\left|\mathbf{a}_{i}\right|$ are areas of the external
(boundary) triangles, and $\theta$ is one of the six 3D dihedral
angles between triangles, while $\boldsymbol{\alpha}$ are four degrees
of freedom we are going to coarse-grain.

Using the analogy of the (2+1) case (but it is not really obvious
since it would require a 4-dimensional space to imagine this object),
the four tetrahedra of the discretized space and the tetrahedron of
the total volume can be arranged so that they construct a flat \textit{4-simplex}:
five 3-simplices (tetrahedra), connected to each other, form a 4-simplex.
This flat 4-simplex is a portion of $\mathbb{R}^{4}$. The four tetrahedra
with the total tetrahedron is the boundary of the 4-simplex and is
automatically embedded on $\mathbb{R}^{4}$. Therefore, we could describe
these tetrahedra vectorially, using a 3-form $\mathbf{v}_{i}\in\Omega^{3}\left(\mathbb{R}^{4}\right),$
such that their norms gives the same norms $\left|\mathbf{v}_{i}\right|$
as before.

By the same reasoning with the (2+1)-dimensional case, we have the
closure constraint or the Minkowski theorem in $\mathbb{R}^{4}$:
\begin{equation}
\sum_{i=1}^{5}\mathbf{v}_{i}=0,\label{eq:21}
\end{equation}
see FIG. 14.

\begin{figure}[h]
\centerline{\includegraphics[scale=0.8]{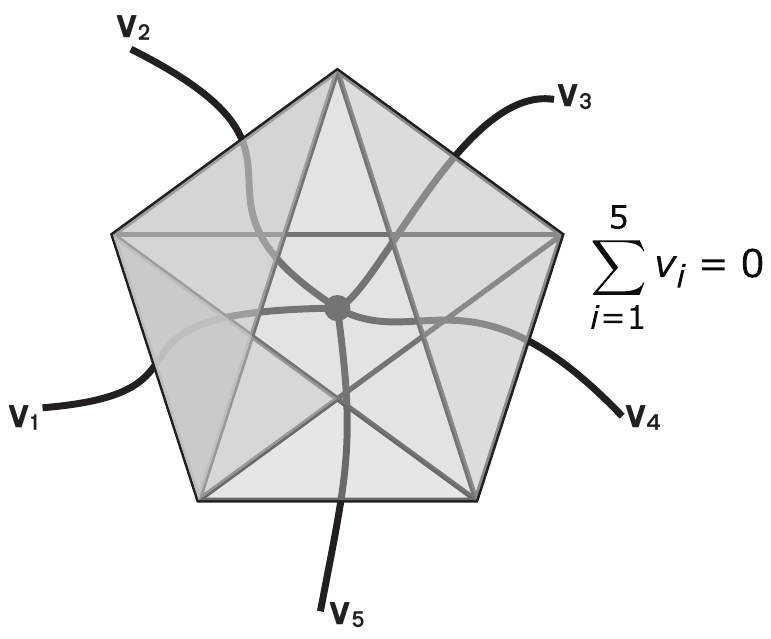}} \caption{The Minkowski theorem in $\mathbb{R}^{4}$, a closed boundary (hypersurface)
of a 4-simplex must satisfy the closure constraint.}
\end{figure}

The gauge invariance condition (\ref{eq:21}) can be written as: 
\[
\mathbf{v}_{1234}=\mathbf{v}_{5}=-\left(\mathbf{v}_{1}+\mathbf{v}_{2}+\mathbf{v}_{3}+\mathbf{v}_{4}\right),
\]
therefore, taking the norm, we obtain the formula relating the total
volume with the fine-grained volumes: 
\begin{eqnarray}
\left|\mathbf{v}_{1234}\right| & = & \sqrt{\sum_{i=1}^{4}\left|\mathbf{v}_{i}\right|^{2}-2\sum_{\binom{i<j}{i=1}}^{4}\left|\mathbf{v}_{i}\right|\left|\mathbf{v}_{j}\right|\cos\varphi_{ij}},\label{eq:23}
\end{eqnarray}
with $\varphi_{ij}$ is the \textit{4-dimensional angle}: the angle
between two tetrahedra of a 4-simplex (this 'angle between spaces'
can only exist in spaces with dimension higher than three), see \cite{key-7}.

Remarkably, this angle also satisfy the dihedral angle formula in
one-dimension higher \cite{key-3.28}: 
\begin{equation}
\cos\varphi_{ij,k}=\frac{\cos\theta_{ij,k}-\cos\theta_{il,k}\cos\theta_{lj,k}}{\sin\theta_{il,k}\sin\theta_{lj,k}},\label{eq:24}
\end{equation}
with $\theta_{ij,k}$ is the dihedral angle between two triangles
$i,j$, located on hinge $k$ \cite{key-7,key-3.28}. Formula (\ref{eq:23})
is the 3D analog to (\ref{eq:4.1}) and (\ref{eq:16}), it is also
purely intrinsic to the 3D space and does not depend on the embedding
in $\mathbb{R}^{4}.$ Using the inverse of (\ref{eq:24}), we could
obtain $\theta_{ij}$, while having the information of $\left\{ \left|\mathbf{v}_{i}\right|,\varphi_{ij}\right\} $,
we could obtain the external boundary triangles $\left|\mathbf{a}_{i}\right|$.

Now let us define the terminologies; the \textit{correct microstate}
is $\left\{ \left|\mathbf{v}_{1234}\right|,\left|\mathbf{a}_{i}\right|,\theta,\boldsymbol{\alpha}\right\} ,$
for $i=1,..,4$, the \textit{$k^{th}$-microstates} are $\left\{ \left|\mathbf{v}_{1234}^{\left(k\right)}\right|,\left|\mathbf{a}_{i}^{\left(k\right)}\right|,\theta^{\left(k\right)},\boldsymbol{\alpha}^{\left(k\right)}\right\} ,$
satisfying:

\begin{equation}
\left|\mathbf{v}_{1234}^{\left(k\right)}\right|=\sqrt{\sum_{i=1}^{4}\left|\mathbf{v}_{i}^{\left(k\right)}\right|^{2}-2\sum_{\binom{i<j}{i=1}}^{4}\left|\mathbf{v}_{i}^{\left(k\right)}\right|\left|\mathbf{v}_{j}^{\left(k\right)}\right|\cos\varphi_{ij}^{\left(k\right)}.}\label{eq:25}
\end{equation}
The \textit{correct reduced-microstate} is $\left\{ \left|\mathbf{v}_{1234}\right|,\left|\mathbf{a}_{i}\right|,\theta\right\} $
, and the \textit{$k^{th}$-reduced-microstates} are $\left\{ \left|\mathbf{v}_{1234}^{\left(k\right)}\right|,\right.$
$\left|\mathbf{a}_{i}^{\left(k\right)}\right|,$ $\left.\theta^{\left(k\right)}\right\} .$
The \textit{coarse-grained tetrahedron} is defined as the \textit{macrostate}
of the system $\left\{ \overline{\left|\mathbf{V}\right|},\overline{\left|\mathbf{A}_{i}\right|},\bar{\theta}\right\} $
: 
\begin{eqnarray*}
\overline{\left|\mathbf{V}\right|} & = & \sum_{k}p_{k}\left|\mathbf{v}_{1234}^{\left(k\right)}\right|,\\
\overline{\left|\mathbf{A}_{i}\right|} & = & \sum_{k}p_{k}\left|\mathbf{a}_{i}^{\left(k\right)}\right|\\
\bar{\theta} & = & \sum_{k}p_{k}\theta^{\left(k\right)},
\end{eqnarray*}
which is an average value over all possible reduced-microstates. The
\textit{coarse-grained volume} is $\overline{\left|\mathbf{V}\right|}$
, while the\textit{ fine-grained volume} is $\sum_{i=1}^{3}\left|\mathbf{v}_{i}\right|:$
the volume of the 'correct' microstate. The microcanonical case is
described by the constraint: 
\[
\left|\mathbf{v}_{1234}^{\left(k\right)}\right|=\left|\mathbf{v}_{1234}\right|.
\]
For the canonical and grandcanonical case, $\left|\mathbf{v}_{1234}^{\left(k\right)}\right|=\left|\sum_{i=1}^{4}\mathbf{v}_{i}^{\left(k\right)}\right|$
and $\left|\mathbf{v}_{1234}^{\left(k\right)}\right|=\left|\sum_{i=1}^{n_{k}}\mathbf{v}_{i}^{\left(k\right)}\right|$,
respectively. The volume and particle number fluctuations are defined
in the same way as in the previous cases. We still have freedom to
coarse-grain length \textit{and} area just as in the previous sections,
but it must be done in order: first coarse-grain the volume, then
the area, and lastly, the length. The reason is because coarse-graining,
in a sloppy sense, is 'adding' $n$-simplices, which are $n-$forms,
and in order to add forms, they need to be embedded on a same (cotangent)
space first.

The 4-1 Pachner move is simultaneously removing four hinges (segments
in (3+1) theory) of the discretized curved space, not only removing
a single hinge as in the (2+1) case. What if we only want to coarse-grained
single hinge in the (3+1) case? This is the 3-2 Pachner move.

\paragraph{3-2 Pachner moves case.}

The simplest coarse-graining case which coarse-grain only a single
hinge, is the 3-2 Pachner move. See FIG. 15. 
\begin{figure}[h]
\centerline{\includegraphics[scale=0.85]{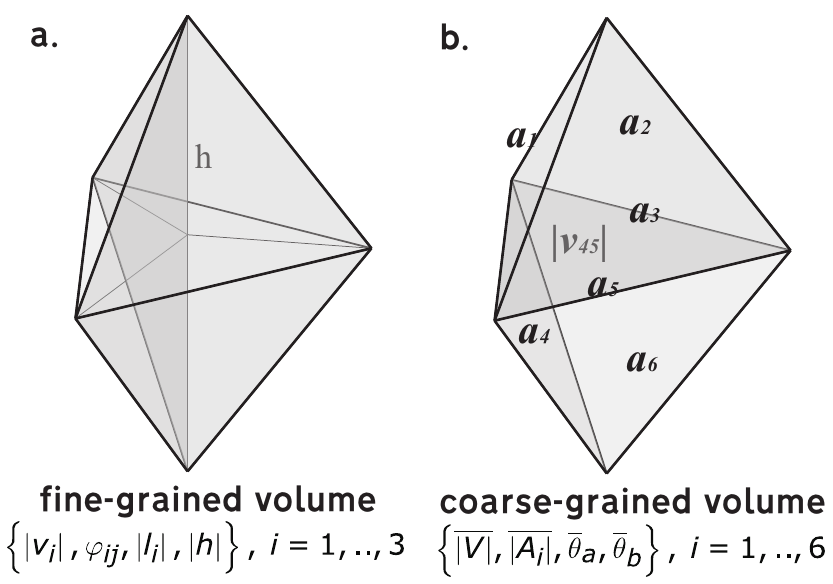}} \caption{(a) is a portion of a discretized space triangulated by three tetrahedra
$\left\{ a,b,c\right\} $. The fine grained state contains three tetrahedra
glued together on their triangles to contruct a \textit{bipyramid}
- a 'diamond' in figure (b). The curvature is located on the hinge
$\mathbf{h}$ inside it. When we coarse-grained, we loose the hinge,
and the coarse-grained bipyramid is now constructed by only two tetrahedra.
This transformation is known as 3-2 Pachner move.}
\end{figure}
 FIG. 15(a), which is the '3' part of the 3-2 Pachner move, has ten
degrees of freedom, this can be easily obtained by calculating the
number of segments of this geometrical object. But instead of using
these ten segments as the coordinate-free variables, to prevent ambiguities,
we use $\left\{ \left|\mathbf{v}_{i}\right|,\varphi_{ij},\left|\mathbf{l}_{i}\right|,\left|\mathbf{h}\right|\right\} $,
for $i\neq j=1,2,3$; with $\left|\mathbf{v}_{i}\right|$ is the volume
of the individual tetrahedron, $\varphi_{ij}$ is the coupling between
each two tetrahedra, $\left|\mathbf{l}_{i}\right|$ is one segment
of each tetrahedron, and $\left|\mathbf{h}\right|$ is the common
internal segment shared by these three tetrahedra. See FIG. 16(a).
\begin{figure}[h]
\centering{}\includegraphics[scale=0.85]{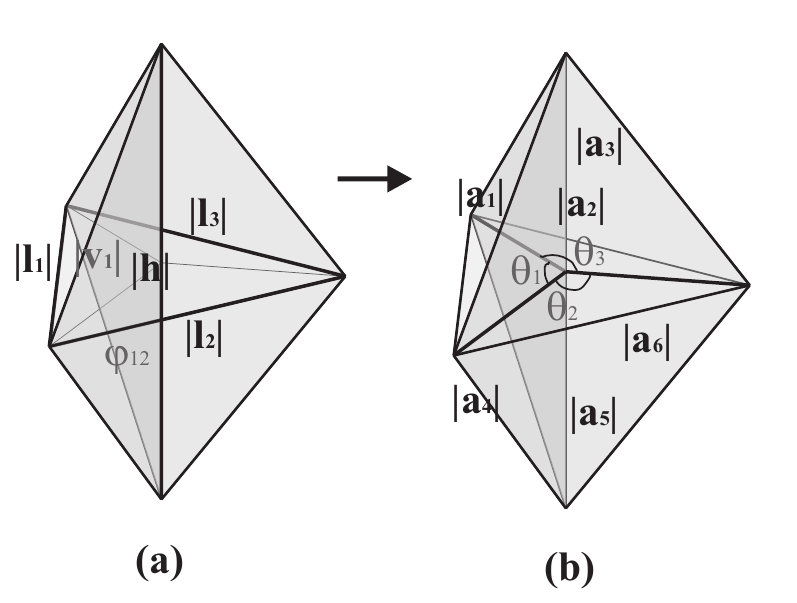}\caption{(a) is the '3' part of the 3-1 move, described by ten variables. (b)
is the '2' part of the 3-1 move, described by nine variables.}
\end{figure}

To coarse-grain FIG. 16(a) means we treat the three coupled tetrahedra
as a single system with the surface boundary is held to be constant
for a moment. This mean, we kept the nine external segments constructing
the 'bipyramid' to be constants, and we remove the internal segment
$\left|\mathbf{h}\right|$ away. But $\left|\mathbf{h}\right|$ is
the location of the intrinsic 3D curvature, that is, it defines the
dihedral angles of each tetrahedron, say, $\left\{ \theta_{1},\theta_{2},\theta_{3}\right\} ,$
which gives the 3D intrinsic curvature:
\[
\delta\theta=2\pi-\sum_{i=1}^{3}\theta_{i}.
\]
Therefore, removing $\left|\mathbf{h}\right|$ away will result in
the lost of the intrinsic 3D curvature. This condition can be realized
by setting $\delta\theta=0,$ so that it is equivalent with introducing
a new constraint:
\begin{equation}
\theta_{1}+\theta_{2}+\theta_{3}=0\label{eq:zerocurve}
\end{equation}
to the system. With this constraint, the total degrees of freedom
of the system in FIG. 16(a) reduces from ten to nine (the '2' part
of the move), which is exactly the number of segments of a flat, trihedral-bipyramid.
We choose the 'center-of-mass' variables to be $\left\{ \left|\mathbf{v}_{123}\right|,\left|\mathbf{a}_{i}\right|,\theta_{j}\right\} ,$
for $i=1,..,6,$ and $j=1,..,3,$ with $\left|\mathbf{v}_{123}\right|$
is the total volume of the 'center-of-mass' bipyramid, satisfying:
\begin{eqnarray*}
\left|\mathbf{v}_{123}\right| & = & \sqrt{\sum_{i=1}^{3}\left|\mathbf{v}_{i}\right|^{2}-2\sum_{\binom{i<j}{i=1}}^{3}\left|\mathbf{v}_{i}\right|\left|\mathbf{v}_{j}\right|\cos\varphi_{ij}},\; i,j=1,2,3.
\end{eqnarray*}
$\left|\mathbf{a}_{i}\right|$ are the six external areas of the bipyramid
triangles, and $\theta_{j}$ are the three dihedral angles located
on the hinge, they determine a segment of a triangle. See FIG. 16(b).

Now let us define the terminologies. Let $\left\{ \left|\mathbf{v}_{123}\right|,\left|\mathbf{a}_{i}\right|,\theta_{j}\right\} $
for $i=1,..,6,$ and $j=1,..,3,$ be the \textit{correct microstate},
then $\left\{ \left|\mathbf{v}_{123}^{\left(k\right)}\right|,\left|\mathbf{a}_{i}^{\left(k\right)}\right|,\right.$
$\left.\theta_{j}^{\left(k\right)}\right\} $ are the \textit{$k^{th}$-microstate}.
By inserting constraint (\ref{eq:zerocurve}) to the system, the \textit{correct
reduced-microstate} is $\left\{ \left|\mathbf{v}_{123}\right|,\left|\mathbf{a}_{i}\right|,\theta_{1},\theta_{2}\right\} $
and the $k^{th}$\textit{-reduced-microstates} are $\left\{ \left|\mathbf{v}_{123}^{\left(k\right)}\right|,\left|\mathbf{a}_{i}^{\left(k\right)}\right|,\theta_{1}^{\left(k\right)},\theta_{2}^{\left(k\right)}\right\} $.
See FIG. 16(b). The \textit{macrostate} is $\left\{ \overline{\left|\mathbf{V}\right|},\overline{\left|\mathbf{A}_{i}\right|},\bar{\theta}_{a},\bar{\theta}_{b}\right\} $:
\begin{eqnarray*}
\overline{\left|\mathbf{V}\right|} & = & \sum_{k}p_{k}\left|\mathbf{v}_{123}^{\left(k\right)}\right|,\\
\overline{\left|\mathbf{A}_{i}\right|} & = & \sum_{k}p_{k}\left|\mathbf{a}_{i}^{\left(k\right)}\right|,\\
\bar{\theta}_{a,b} & = & \sum_{k}p_{k}\theta_{a,b}^{\left(k\right)},
\end{eqnarray*}
containing nine degrees of freedom. We can derive all the same class
of properties as in the 4-1 Pachner move case.

\section{Discussion}

Following the classical coarse-graining procedures we have already
proposed in the main part of this article, we will discuss four related
subject which could be the first steps to understand the problems
mentioned in the Introduction: the thermodynamical properties of general
relativity and the correct classical limit of quantum gravity. These
subjects include an extension to a non-compact space case, the inverse
procedure of coarse-graining called as refinement, and limits in statistical
mechanics.

\subsection{Non-compact space and infinity}

Let us recall our first and second restrictions, respectively: (1)
the sample space must be discrete, and (2) the sample space must be
finite. It should be kept in mind that the second restriction is only
applied to a finite system, which in our case, a compact spatial space.
These two restrictions guarantee that given a finite system, we have
a finite sample space containing finitely all possible microstates
of the finite system. This is important because according to the atomism
philosophy, it is necessary to have a finite amount of informations
given a finite system, so that we only need to provide finite informations
to know exactly the 'correct' microstate of the system. 

Let us recall a bit of the atomism theory. By definition, the 'atoms'
of any entity can not be arbitrarily small, nor arbitrarily large.
If they are arbitrarily small, we could have continuous physical entities
which is forbidden in atomism, while if they are arbitrarily large,
we could have infinitely large, undivided physical entities, which
is also in contrast with atomism. Atomism allow us to reduce finite
(or infinite) things into finitely (or infinitely) many parts with
finite size. These finite parts are atoms. Inversely, we obtain finite
things by adding many atoms finitely, and obtain infinite things by
adding atoms infinitely. The point is, whether the things is finite
or infinite, they can be broken down into countable minimal parts.

In the previous part of our work, the space foliation of spacetime
where we applied coarse-graining is always taken to be compact, say,
two-coupled edges, three coupled triangles, three and four coupled
tetrahedra. Now, we generalize the case such that the foliation is
a \textit{non-compact} space. A non-compact space can be thought of
as an infinite \textit{connected sum} of compact spaces with boundary,
that is, by patching their boundary together: 
\[
\overset{\infty}{\#}M_{\textrm{compact}}=M_{\textrm{noncompact.}}
\]
Using this definition and applying the cut-offs to the non-compact
foliation case, the first restriction is still valid, so $S_{\textrm{min}}=0.$
But the second restriction is \textit{not} valid in the non-compact
case (this is the reason of calling the first restriction as 'strong'
and second restriction as 'weak', since the first is valid for both
compact and non-compact cases, while the later is only valid for compact
case). Therefore, the sample space of the non-compact case is not
finite, and there is no upper bound on its informational entropy:
$S_{\textrm{max}}=\infty.$ But this infinity (the infrared one) is
'acceptable' since it comes from the fact that we are considering
an infinite entity: the non-compact space itself.

The important point is, independent of the compactness/ non-compactness
of the space, these cut-offs maintain the \textit{atomism} point of
view as a foundation in this theory. Space is constructed from the
\textit{atoms} of space: a set of finite elements of building-blocks
having various finite size (length, areas, volume), within the range
described by (\ref{eq:discrete}).

Atomism is a way to 'finitize' entities \cite{key-12}. It is necessary
to have the 'atoms' not to be infinitely small, nor infinitely large.
It is the 'quantum': a discrete entity \cite{key-13}.

\subsection{Refinement}

We define the \textit{refinement} map as the inverse map of coarse-graining.
In contrast with the coarse-graining map, we need to \textit{provide}
information for each step of the refining map. For an example, to
describe completely the motion of each part of a many-body problem,
we need to provide informations of the individual bodies. Because
of the existence of minimal scale in nature, there exist an upperbound
for refinement, this minimal scale act as a physical cut-off which
prevent the UV-divergencies. The upperbound on the refinement is in
accordance with the statement that the amount of information in the
universe is finite \cite{key-2.7}.

\subsection{Continuum \textit{v/s} thermodynamical limit}

There are two types of \textit{limits} in statistical mechanics \cite{key-15,key-15a}:
the \textit{continuum limit} and the \textit{thermodynamical limit},
both are taking the number of partitions to be large: $n\rightarrow\infty,$
but in different manners. The \textbf{continuum limit} is obtained
by taking $n\rightarrow\infty,$ with additional requirements: (1)
all microscopic and intensive quantities (the individual length, area,
and volume $\left\{ \left|\mathbf{l}_{i}\right|,\left|\mathbf{a}_{i}\right|,\left|\mathbf{v}_{i}\right|\right\} $
in this case) becomes arbitrarily small, they go to zero; and (2)
all macroscopic extensive quantities (the total length, area, and
volume of the system $\left\{ \left|\mathbf{L}\right|,\left|\mathbf{A}\right|,\left|\mathbf{V}\right|\right\} $)
are constants, or at least, asymptotically constant. Meanwhile, the
\textbf{thermodynamical limit} is obtained also by taking $n\rightarrow\infty,$
but with additional requirements: (1) all microscopic and intensive
quantities are constant, or at least asymptotically. (2) all macroscopic
extensive quantities \textit{increase} with the number of partitions
$n.$ This means $\left\{ \left|\mathbf{L}\right|,\left|\mathbf{A}\right|,\left|\mathbf{V}\right|\right\} \rightarrow\infty,$
in this limit. It had been shown that both of these limits are equivalent
classically \cite{key-15}, but it is not clear if this is also the
case in general, particularly, for a background independence theory. 

Let us implement these limits to the statistical discrete geometry
description. If we take the continuum limit, the first restriction
will be violated, since the size of partition will become arbitrarily
small, which is forbidden by the first restriction. This will lead
to the UV-divergence in the level of entropy.

Other contradiction which occurs if we take the continuum limit is
related to the background independence point of view. The total size
(total length, total area, total volume) of the space can not be fixed
as $n\rightarrow\infty,$ because of the background independence.
The quanta of space \textit{is} the space itself, it is \textit{not}
atoms which lives in space so that we can add more and more of them
on a fixed background space. Therefore, we argue that the continuum
limit is inconsistent with the atomism and background independence
philosophy.

On the contrary, if we take the thermodynamical limit, it will violate
the second restriction for a compact case (which is acceptable, since
the second restriction is only valid for compact case%
\footnote{According to the atomism philosophy, a compact space is not compatible
with the $n\rightarrow\infty$ limit: if we add finite entities infinitely,
we should obtain infinite entity. Therefore, a compact space can only
have finite number of partition.%
}). But this is \textit{not} a problem for a non-compact case, since
the size of the non-compact space is infinite and it needs only to
satisfy the first restriction. In the other hand, for the thermodynamical
limit, when we take $n\rightarrow\infty,$ the extensive macroscopic
variables $\left\{ \left|\mathbf{L}\right|,\left|\mathbf{A}\right|,\left|\mathbf{V}\right|\right\} \rightarrow\infty,$
which is consistent with the background independence concept.

Finally, we argue that the correct/ consistent limit of the (Regge)
discrete geometry obtained by the refinement procedure it \textit{not}
the continuum limit, but the thermodynamical limit. The existence
of this limit will cause the macroscopic variables pretend \textit{as
if} they are resulting from a continuous theory \cite{key-15b}, which
might be an effective theory for a large $n$ discrete geometry. The
macroscopic variables of this effective theory are related by the
equation of state, where they satisfy the laws of thermodynamics.
The next questions are: \textit{Is the equation of state resulting
from this thermodynamical limit compatible with general relativity?
Can general relativity be derived from an equation of state?} Some
studies suggest that for a special case, general relativity could
be derived from an equation of state, providing the entropy is proportional
to the area $S=\frac{A}{4}$ \cite{key-16}, but of course this is
another story.

\subsection{Quantum analogy?}

The general idea is already presented in \cite{key-4.3} (in fact,
this article is the classical analogy which is based by the formulation
in \cite{key-4.3}), based on the canonical loop quantum gravity theory,
but it will need a more careful, detailed, and precise explanation.

\subsection{Conclusion}

We have construct the statistical discrete geometry by applying statistical
mechanics to discrete (Regge) geometry. We have propose an averaging/coarse-graining
method for discrete geometry by maintaining two philosophical assumptions:
atomism and background independence concept. To maintain atomism and
background independence philosophy, we propose restrictions to the
theory by introducing cut-offs, both in ultraviolet and infrared regime.
In discrete geometry, this cut-offs truncate the theory with infinite
degrees of freedom into a theory with finite degrees of freedom. The
interaction between two partitions (quanta) of geometries manifest
through the intrinsic curvature, as a 'coupling' in the theory. Using
the infinite degrees of freedom limit, we argue that the correct limit
consistent with the restrictions and the background independence concept
is not the continuum limit of statistical mechanics, but the thermodynamical
limit. If this thermodynamical limit exist, theoretically, we could
obtain the corresponding equation of states of statistical discrete
geometry, which is expected to be general relativity. Works to find
this limit is highly encouraged, it might be the first step to understand
the thermodynamical aspect of general relativity. The quantum version
of this article is under progress.

   \bibliography{library}
	
	\end{document}